%% file: paper_re.tex
\documentstyle[prd,12pt,epsfig,aps]{revtex}

\def\altaffilmark#1{$^{#1}$}
\def\altaffiltext#1#2{\footnotetext[#1]{#2}\stepcounter{footnote}}

\renewcommand{\thefootnote}{\arabic{footnote}}

\draft

\begin{document}

\narrowtext
\bibliographystyle{prsty}

\title{The Cosmic Ray Proton Spectrum \\
determined with the \\
Imaging Atmospheric Cherenkov-Technique}
\input{authors}
\altaffiltext{1}{Max-Planck-Institut f\"ur Kernphysik,
Saupfercheckweg 1, D-69117 Heidelberg, Germany}
\altaffiltext{2}{Max-Planck-Institut f\"ur Physik, F\"ohringer Ring
6, D-80805 M\"unchen, Germany}
\altaffiltext{3}{Universidad Complutense, Facultad de Ciencias
F\'{\i}sicas, Ciudad Universitaria, E-28040 Madrid, Spain}
\altaffiltext{4}{Universit\"at Hamburg, II. Institut f\"ur
Experimentalphysik, Luruper Chaussee 149,
D-22761 Hamburg, Germany}
\altaffiltext{5}{Universit\"at Kiel, Institut f\"ur Physik,
Leibnitzstr. 15, D-24118 Kiel, Germany}
\altaffiltext{6}{Universit\"at Wuppertal, Fachbereich Physik,
Gau{\ss}str. 20, D-42097 Wuppertal, Germany}
\altaffiltext{7}{Yerevan Physics Institute, Yerevan, Armenia}
\altaffiltext{8}{Now at Department of Physics University of Leeds,
Leeds LJ2 9JT, UK}
\altaffiltext{9}{Now at Forschungszentrum Karlsruhe, P.O. Box 3640, 76021
Karlsruhe, Germany}
\altaffiltext{10}{On leave from Altai State University, Barnaul, Russia}
\altaffiltext{11}{Now at SAP AG, Neurottstr. 16, D-69190 Walldorf, Germany}
\altaffiltext{12}{Now at Universidad Aut\'{o}noma de Barcelona, Institut de
F\'{\i}sica d'Altes Energies, E-08193 Bellaterra, Spain}
\altaffiltext{13}{Now at Enrico Fermi Institute, The University of Chicago,
933 East 56$^{\rm th}$ Street, Chicago, IL, 60637, USA}

\renewcommand{\thefootnote}{\arabic{footnote}}

\date{\today}

\maketitle

\begin{abstract}
   The HEGRA system of 4 Imaging Atmospheric Cherenkov Telescopes (IACTs)
   has been used to determine the flux and the spectrum of
   cosmic ray protons over a limited energy range around 1.5~TeV.
   Although the IACT system is designed for the detection of
   $\gamma$-rays with energies above 500~GeV, it has also a large detection
   area of $\,\simeq$\,10$^6$~m$^2$ $\cdot$ 3~msr
   for primary protons of energies above 1~TeV
   and the capability to reconstruct the primary proton energy with
   a reasonable accuracy $\Delta$E/E of 50\% near this threshold.
   Furthermore, the principle of stereoscopic detection of air showers
   permits the effective suppression of air showers induced by heavier
   primaries already on the trigger level, and in
   addition on the software level by analysis of the stereoscopic
   images. The combination of both capabilities permits a determination of
   the proton spectrum almost independently of the cosmic ray chemical
   composition. The accuracy of our estimate of the spectral index at 1.5~TeV
   is limited by systematic uncertainties and is comparable to the accuracy
   achieved with recent balloon and space borne experiments. In this paper we
   describe in detail the analysis tools, namely the detailed Monte Carlo
   simulation, the analysis procedure and the results. We determine the local
   (i.e.~in the range of 1.5 to 3 TeV) differential spectral index to be
   $\gamma_p = 2.72\pm 0.02_{\rm stat.}{\pm 0.15}_{\rm syst.}$ and obtain an
   integral flux above 1.5~TeV of $F(\,> 1.5 {\rm\, TeV}) =3.1\pm 0.6_{\rm
   stat.}\pm 1.2_{\rm syst.}\cdot 10^{-2}/$s sr m$^2$.
\end{abstract}
\newpage

%
%***************************************************************************
   \section{Introduction}
      %*****************************************************************
      The stereoscopic system of Imaging Atmospheric Cherenkov telescopes
      (IACT-system) of the HEGRA collaboration \cite{DAU97}
      is a powerful tool for detecting TeV $\gamma$-ray sources and for
      performing detailed spectroscopic studies in the energy range from
      500~GeV to $\sim$ 50~TeV, where the latter limit is determined by
      event statistics alone.
      With the nearly background-free detection of $\gamma$-rays from the
      Crab Nebula \cite{DAU97}, an
      energy flux sensitivity $\nu\,F_\nu$ of $\simeq$\,10$^{-11}$
      ergs/cm$^2$ s at 1 TeV for one hour of observation time has been
      estimated. The high signal to noise ratio together with the energy
      resolution of better than 20\% for primary photons makes it possible to
      study the spectra of strong sources on time scales of one hour, as
      demonstrated by the observation of the BL Lac object Mkn 501
      during its 1997 state of high and variable emission
      \cite{AHA98}.

      The IACT system can not only be used for $\gamma$-ray
      astronomy. It can also contribute to the
      study of charged cosmic rays (CR) for energies between a few TeV and
      possibly $\sim$\,100
      TeV, a key energy region for the understanding of
      the sources of CRs and their propagation
      through our galaxy (see e.~g.~\cite{SWO93,SHI95} and references therein
      for reviews).

      The measurement
      with the IACT system described in this paper has systematic
      uncertainties comparable to recent measurements of  satellite and
      balloon borne experiments (see e.~g.~\cite{WIE97} for a recent
      compilation). A clear advantage of the IACT technique is the large
      effective area of $\simeq$\,3$\cdot$10$^3$~m$^2$~sr for TeV cosmic rays
      combined with a field of view of $\simeq$ 3 msr,  corresponding to a
      detection rate of around 12~Hz for $\,>\,$1~TeV cosmic rays.

      In an earlier paper \cite{PLY98} we explored the possibilities
      to use the IACT technique to measure the energy spectra and mass
      composition of CRs and especially CR protons.
      The stereoscopic observation of air showers with at least two IACTs
      suppresses heavier primaries already on the trigger level. This is
      because the energy threshold E$_{\rm thr}$, defined as the energy where
      the differential detection rate peaks, increases substantially with the
      nucleon number $A$, approximately as E$_{\rm thr}\,\propto\,$
      $A^{0.5}$. The {\it stereoscopic} detection of the air shower under
      different viewing angles with high resolution imaging cameras permits
      us to unambiguously reconstruct the air shower axis in three
      dimensions. Knowing the location of the shower core with a precision of
      30~m, the energy of a primary proton can be determined with
      an accuracy $\Delta E/E$ of 50\% and the different projections
      of the longitudinal and lateral shower development
      can be used to obtain an event sample enriched with particles
      of a certain primary species.
      The net effect of the trigger scheme and of the software cuts
      is a suppression of heavier nuclei by a factor larger than 10 at TeV
      energies. This makes the extraction of an almost pure proton data
      sample possible and permits the determination of its energy spectrum,
      at least in a narrow range around 1.5 TeV. Even a rather limited
      knowledge of the CR chemical composition significantly extends this
      dynamical range.

      In this paper we give a detailed description of the principles
      underlying a proton measurement. Then we apply the method
      to data from the HEGRA experiment, that automatically accumulates CR
      air shower data in the form of background events during $\gamma$-ray
      observations. The HEGRA experiment is introduced in Section
      \ref{hegra}, the analysis tools are described in Section \ref{anatool}
      and the results and the systematic errors are presented in Section
      \ref{resu}. Section \ref{disc} discusses the results. This paper is
      based on the results of the PhD thesis \cite{HEM98}.

      %
      %***************************************************************************
   \section{The HEGRA IACT-System}
      %***************************************************************************
      %
      \label{hegra}
      The HEGRA experiment, located on the Canary Island La Palma at the
      Observatorio del Roque de los Muchachos (2200 m a.s.l., 28.75$^\circ$
      N, 17.89$^\circ$ W), is a large detector complex dedicated
      to the study of cosmic rays and $\gamma$-ray astronomy
      \cite{LIN97}. In particular, the HEGRA collaboration operates two air
      shower arrays on a surface of $\simeq4\,\cdot\,10^4$~m$^2$. The first
      one is an array of 243 scintillation detectors of one square meter area
      each \cite{Kraw:96} which samples the particle cascade reaching the
      observation level. The other one is the AIROBICC array of 97 wide angle
      Cherenkov counters \cite{Karl:95} which samples the atmospheric
      Cherenkov photons emitted by the particle cascade.
      Apart from $\gamma$-ray astronomy in the energy range above 15\,TeV,
      the arrays are used to measure the all particle spectrum and the
      chemical composition in the energy range above $\simeq$\,200\,TeV
      \cite{Cort:97,Lind:98}. The third element in operation is the
      stereoscopic IACT system together with two IACTs observing in single
      telescope mode. One of these telescopes has very recently been
      incorporated into the stereoscopic system. Here we concentrate on
      results obtained by the IACT system.

      At the time when the data used in this analysis were taken, the
      stereoscopic IACT-System  consisted of 4 telescopes with 8.5 m$^2$
      mirror area each. Each telescope is equipped with  a 271
      pixel camera, covering a field of view of 4.3$^\circ$. The pixel size
      is 0.25$^\circ$. The cameras are readout by an 8 bit 120 MHz Flash-ADC
      system.

      The telescope system uses a multi level
      trigger scheme \cite{BUL97}.  A coincidence of two neighboring pixels
      above a given threshold triggers an individual telescope.  This trigger
      condition is called $2{\rm NN}/271 > q_0$ hereafter, where NN denotes
      the next-neighbour condition and $q_0$ is the threshold in units of
      registered photoelectrons.
      A coincidence of at least two telescopes (named hereafter 2/M-telescope
      multiplicity, with M=4)  triggers the telescope system and results in
      the readout of the buffered FADC information of all telescopes.

      An absolute calibration of the system has been performed with a
      laser measurement and a calibrated low-power photon detector
      \cite{FRA97a}. This measurement has determined the conversion factor
      from photons to FADC counts with an accuracy of 12\%.
      The error on the energy scale is estimated to be
      15\% which derives from the uncertainty in the conversion factor from
      Cherenkov photon counts to FADC counts, and from the uncertainty in the
      atmospheric absorption.

      To obtain the data used in the following analysis, the photomultipliers
      are operated in a regime where saturation effects due to space charges
      are smaller than 10\%, for less than 400 photoelectrons per pixel. A
      total amplitude of the image, the {\sl Size}, of 400 photoelectrons
      represents an energy of protons of around 15 TeV.

      %
      %***************************************************************************
   \section{Analysis Tools}
      \label{anatool}
      \subsection{Monte Carlo Simulations}
	 %***************************************************************************
	
	 The CR-induced extensive air showers have been simulated with the
	 ALTAI code \cite{PLY89,KON92,PLY98a}.  The simulation of the
	 electromagnetic shower development models the elementary processes
	 of bremsstrahlung, ionization losses and Coulomb scattering of
	 charged particles as well as pair production and Compton scattering
	 of photons. The effect of multiple scattering of the charged
	 particles is simulated with a fast semi-analytical algorithm
	 which computes the probability distributions of the lateral and
	 angular distributions of charged particles
	 in a given volume in space. The simulation of the hadron component
	 is based on accelerator data of
	 $pp$- and $np$-interactions using, where necessary, extrapolations
	 of the cross sections to TeV energies.  The code uses a modified
	 version of the radial-scaling model \cite{HIL79}. Taking into
	 account the probability coefficients for the different fragmentation
	 channels, the model of independent nucleon interactions was used to
	 describe the fragmentation of the nuclear projectile.
	
	 In order to study the model dependence of the observable parameters,
	 a second air shower library was generated, using the CORSIKA code
	 (Version 4.50) \cite{KFK92,KFK98} to simulate the hadronic
	 interactions of the air shower cascade.  CORSIKA offers several
	 interaction models. High energy interactions ($E_{\rm CM}\,>\,80$
	 GeV) were simulated with the HDPM (`hadronic interactions inspired
	 by the Dual Parton Model') \cite{CAP92}.  Low energy interactions
	 ($E_{\rm CM}\,<\,80$ GeV) were modeled with the GHEISHA code (`Gamma
	 Hadron Electron Interaction Shower Algorithm').  HDPM is known to
	 describe, also for heavier primary particles, reasonably well the
	 available accelerator data in the energy region relevant here
	 ($10^{11} < E_{\rm Lab} < 10^{14}$ eV).  Instead of EGS our
	 variant of CORSIKA uses the ALTAI code to model the electromagnetic
	 shower development.
	
\begin{figure}[htbp]
	 \begin{center}
	    \mbox{\epsfig{file=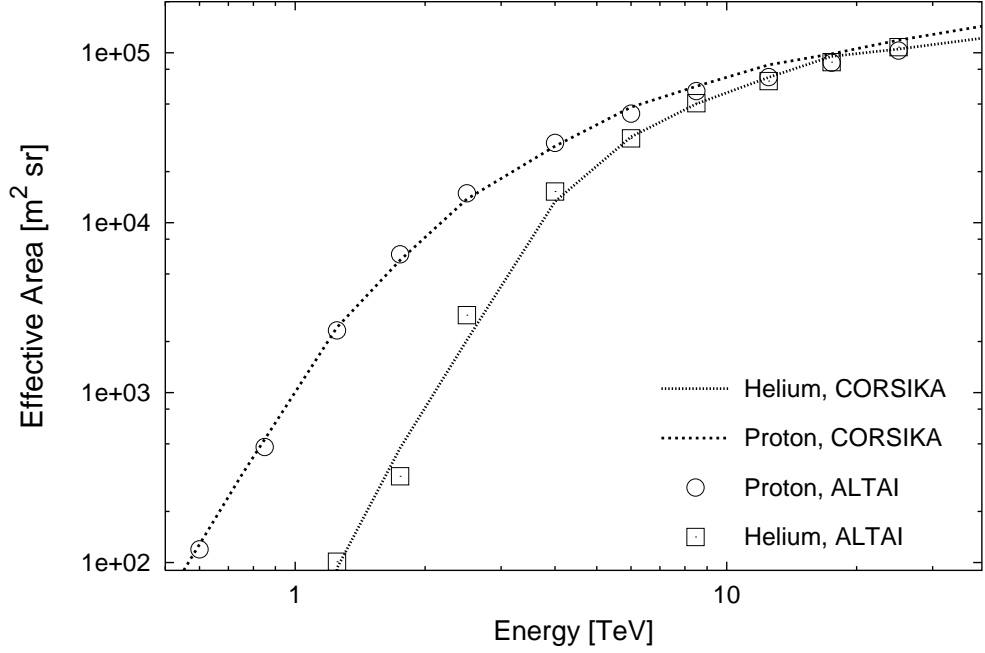,width=0.8\textwidth}}
	 \end{center}
	 \caption{Comparison of effective areas for the System of
	 HEGRA-Cherenkov-telescopes for proton- and helium-induced
	 showers, simulated with the ALTAI hadronic interaction model and
	 the CORSIKA-HDPM code. The differences are smaller than 10\%. The
	 trigger required $2{\rm NN}/271  >10$ ph.e.~and a 2/4 telescope
	 coincidence.}\label{fig-altcor}
      \end{figure}

	 A first comparison of the essential characteristics was performed
	 using $5\cdot 10^5$ showers of vertical incidence, simulated both
	 with the ALTAI and the CORSIKA hadronic interaction models in an
	 energy range of 0.3 to 50 TeV and a distance scale of 250~m to the
	 central telescope of the system. The construction of an energy
	 spectrum relies on the determination of the effective areas. In
	 Figure \ref{fig-altcor} the effective areas for proton- and
	 helium-induced showers, as computed with the two interaction models,
	 are compared with each other. The difference between the two models
	 is smaller than 10\% over the full energy range. Although completely
	 different interaction models have been used, the agreement is
	 excellent. Predicted HEGRA detection rates have been computed,
	 weighting the individual showers according to the chemical
	 composition of the nuclei as known from the literature
	 (\cite{WIE94}, see Table \ref{tab-wiebel}). The predictions of both
	 models, summarized in Table \ref{tab-altcor}, are in very good
	 agreement.
	
\begin{figure}[htbp]
	 \begin{center}
	    \mbox{\epsfig{file=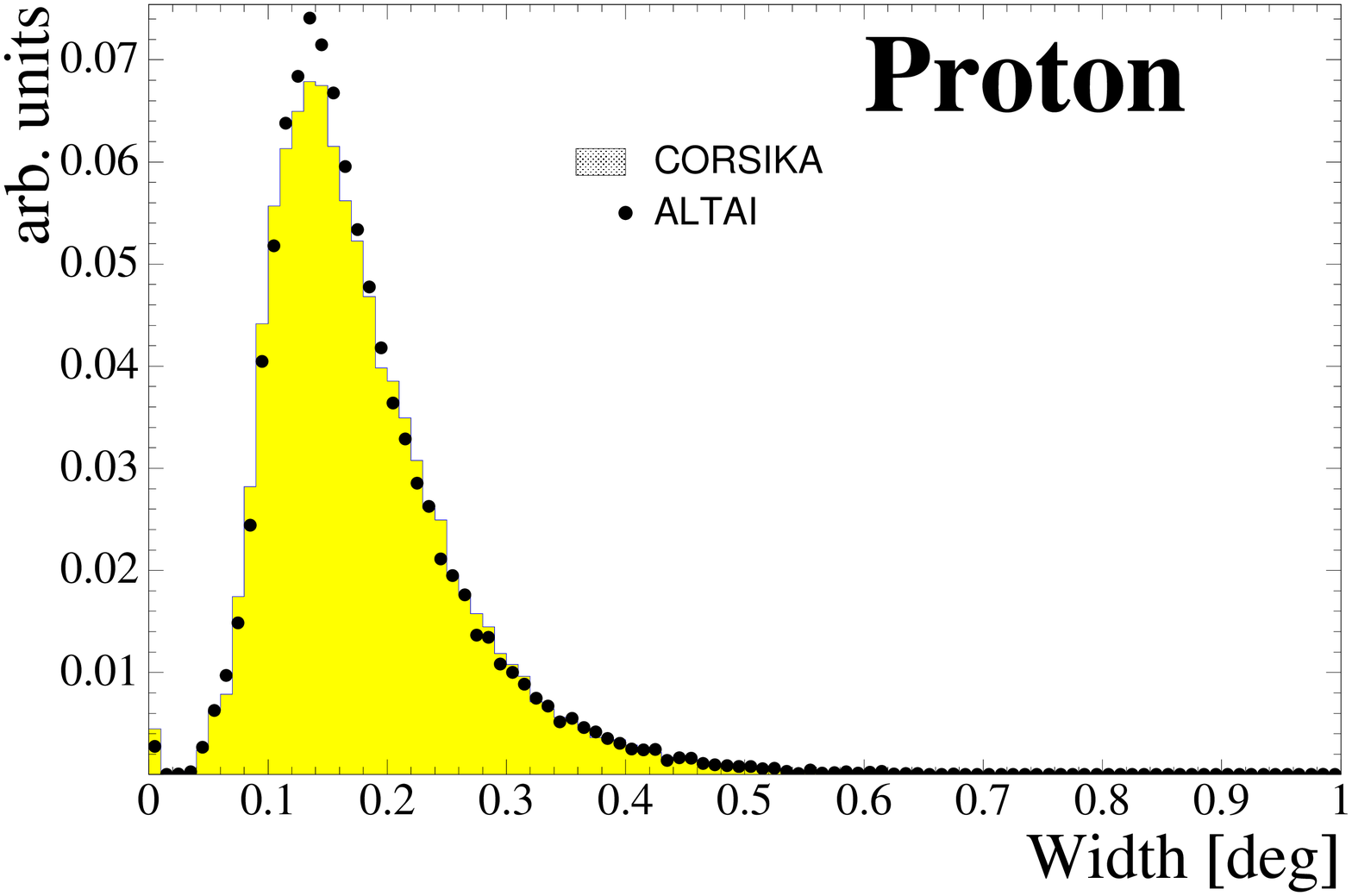,width=0.45\textwidth}}
	    \mbox{\epsfig{file=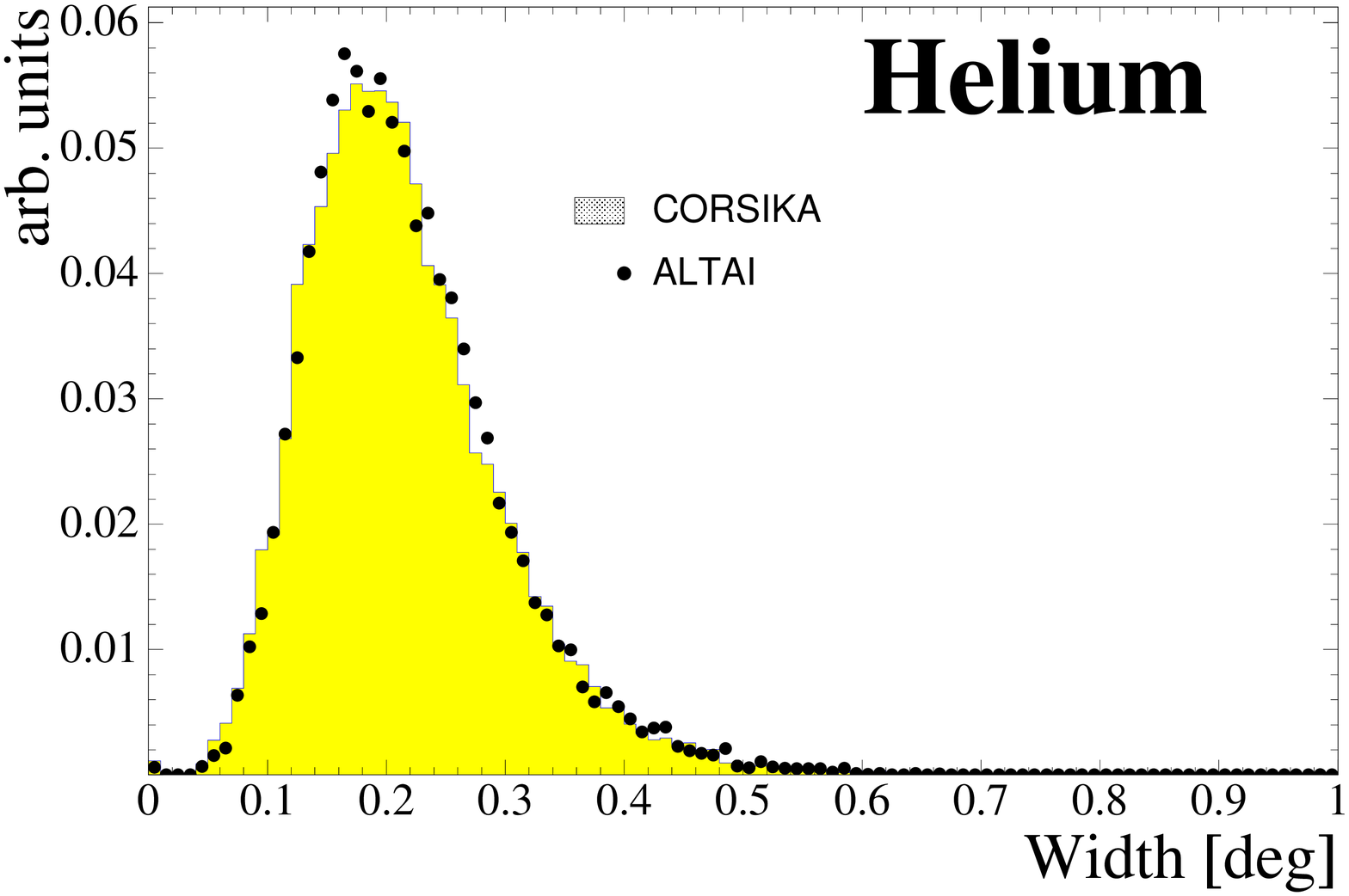,width=0.45\textwidth}}
	    \mbox{\epsfig{file=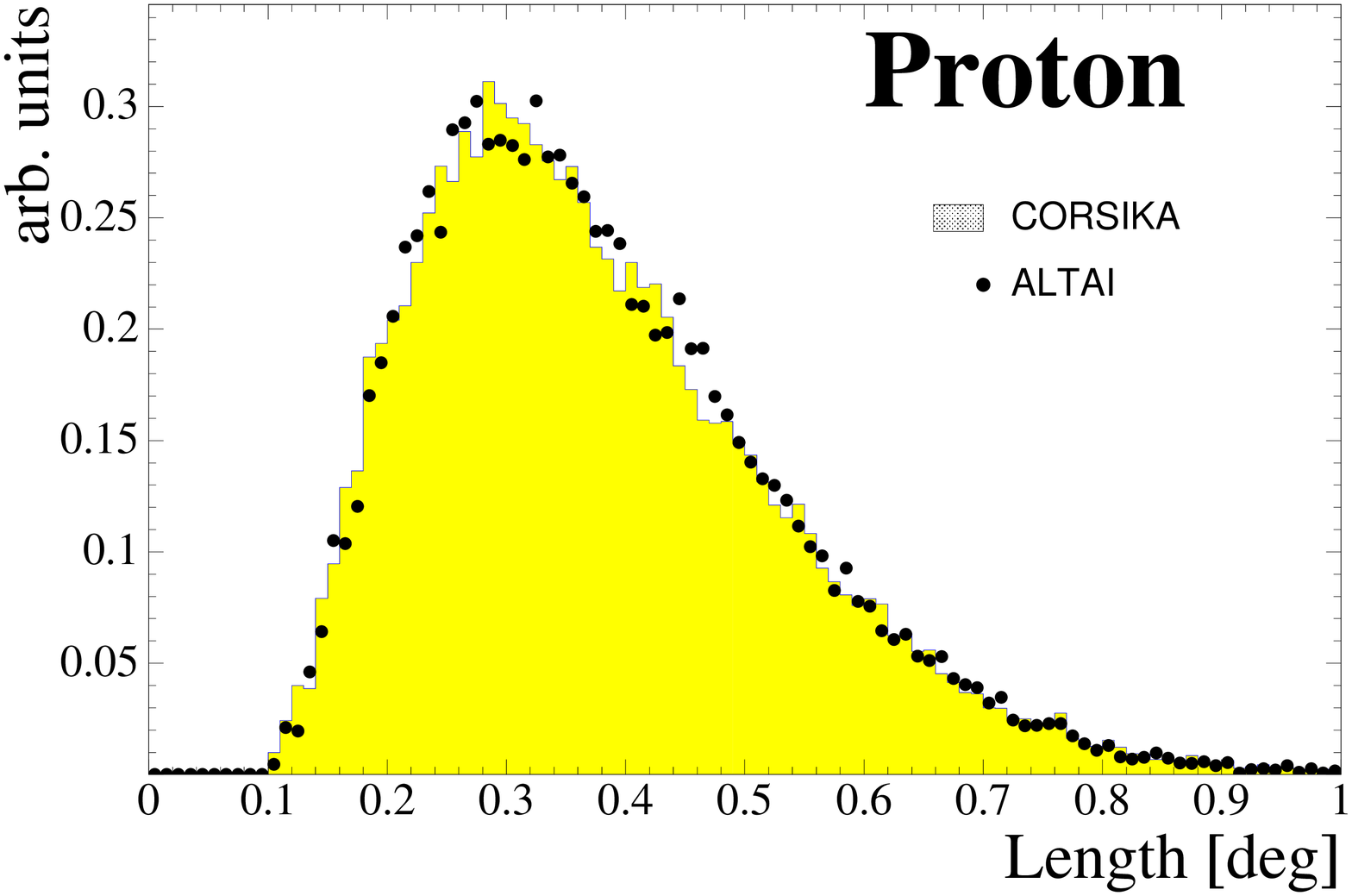,width=0.45\textwidth}}
	    \mbox{\epsfig{file=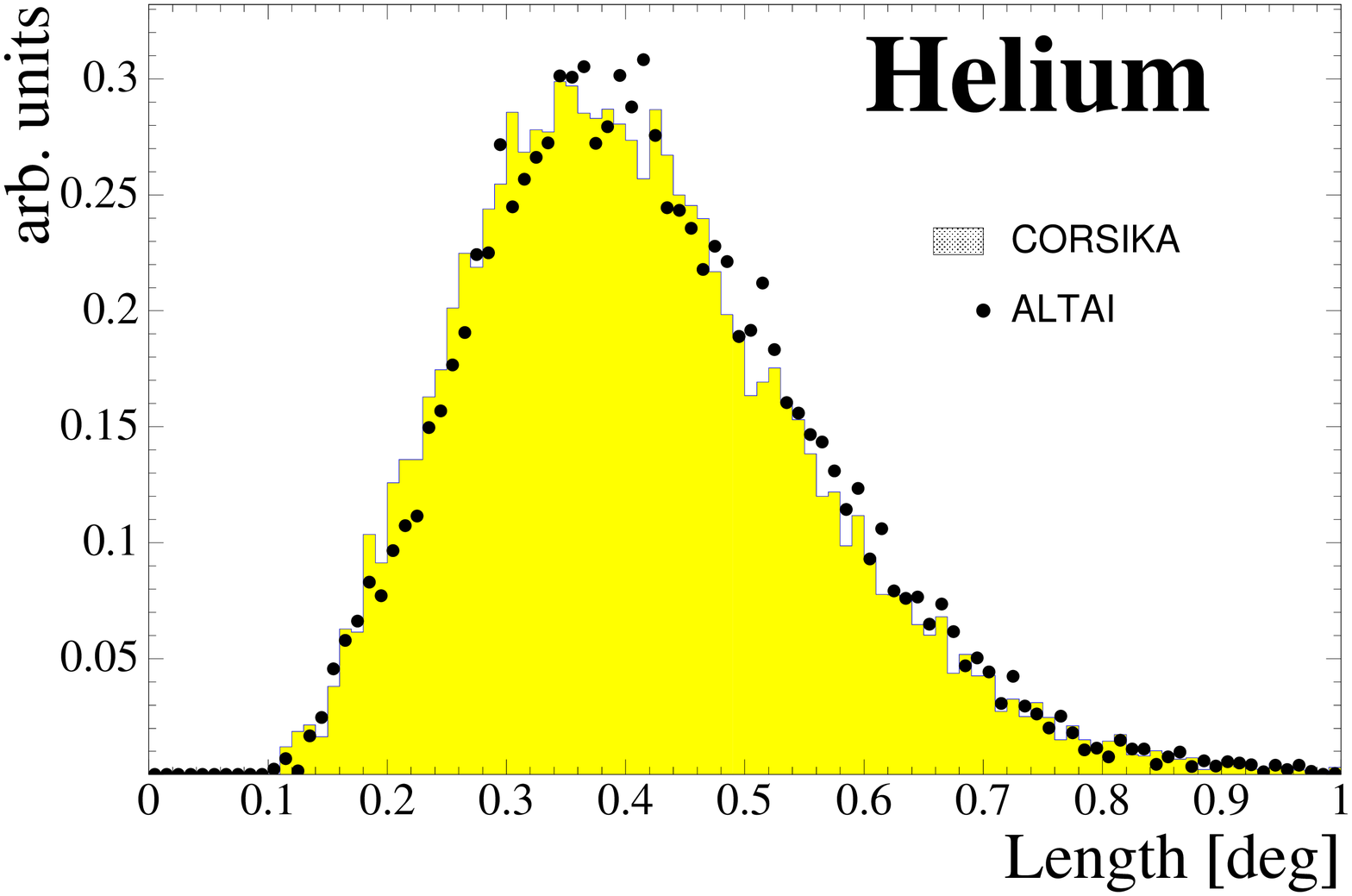,width=0.45\textwidth}}
	 \end{center}
	 \caption{Comparison of the {\sl Width} and {\sl Length} parameter
	 distribution for the proton- and helium-induced showers. The two
	 different interaction models show nearly identical distributions
	 of these image parameters, reflecting the lateral and
	 longitudinal shower development respectively.}\label{fig-img}
      \end{figure}

	 To characterize the telescope images, a 2$^{\rm
	 nd}$-moment analysis is used to derive the standard Hillas
	 parameters \cite{HIL85}, i.e.\
	 the {\sl Width}-parameter which reflects the lateral development of
	 the air shower, and the {\sl Length}-parameter which is related to
	 the longitudinal shower development. Figure \ref{fig-img} compares
	 the {\sl Width} and {\sl Length} parameter distribution as derived
	 with the two interaction models for proton- and helium-induced air
	 showers. The agreement is good.
	
	 In the following a set of $\approx 10^6$ CORSIKA generated showers
	 in the energy and distance range given above is used to analyze the
	 data, comprising simulations for air showers induced by proton and
	 helium as well as by light and medium nuclei (with mass numbers 6 to
	 19, in the following abbreviated with LM), and finally by heavy and
	 very heavy nuclei (with mass numbers 20 to 56, abbreviated as HVH).
	 For the LM- and HVH-groups the atomic numbers of the primary nuclei
	 were randomly distributed inside the group.
	
	 In addition to vertical proton-showers, proton showers incident
	 under $z\,=$ 20$^\circ$ zenith angles were simulated in order to
	 interpolate the effective area for $z\,\in \left[ 0^\circ, 20^\circ
	 \right]$. The effective area varies in this range only weakly according to the
expected $\cos z$-dependence.
	
	 After the air shower simulation, the showers are processed with a
	 new detector simulation of the HEGRA-System of IACTs.  This improved
	 detector simulation includes a full detector simulation, taking into
	 account Cherenkov photon losses due to atmospheric absorption and
	 scattering and due to the telescope mirror, the mirror geometry and
	 the arrival times of the Cherenkov photons, the photomultiplier (PM)
	 response and the characteristics of the electronic chain to derive
	 the trigger decision and the digitized signal.  The new simulations
	 permit an identical treatment of Monte Carlo simulated showers and
	 real data. A detailed description can be found in \cite{HEM98}.
	 %
	 %***************************************************************************
      \subsection{Proton Enrichment of the Data Sample}
	 %***************************************************************************
	
	 The proton component can effectively be separated from heavier
	 cosmic rays over the energy range from 1 to more than 10 TeV
	 \cite{PLY98}.  The suppression of heavier CRs is based on the
	 following air shower characteristics: At a given energy, showers
	 induced by heavier nuclei develop at substantially greater heights
	 in the atmosphere since the cross section $\sigma_A$ for an
	 inelastic hadronic interaction of a primary of nucleon number $A$
	 with the air nuclei increases wit $A$: to first order approximation,
	 $\sigma_A$ is given by $ \sigma_g = \sigma_0 \cdot A ^\alpha$, with
	 $\sigma_g$ being the geometric cross section, $\sigma_0 \approx 30 -
	 50 {\rm\, mb}$, and $\alpha \, \approx \, 2/3$.  In addition, the
	 ratio of transverse momentum to total momentum in the first
	 interaction increases with increasing nucleon number. Also, the
	 momentum of the primary is for heavy primaries shared among several
	 nucleons and the typical transverse momentum generated in
	 interactions is fixed, leading to a larger lateral extension of
	 showers induced by heavy particles.
	 Furthermore, the fraction of energy channeled into electromagnetic
	 subshowers, responsible for the emission of Cherenkov photons,
	 decreases with increasing nucleon number \cite{Lind:98}. The
	 combination of all these effects results in a larger but less
	 intensive Cherenkov light pool, increasing the threshold energy of
	 heavier particles.

\begin{figure}[htbp]
	 \begin{center}
	    \mbox{\epsfig{file=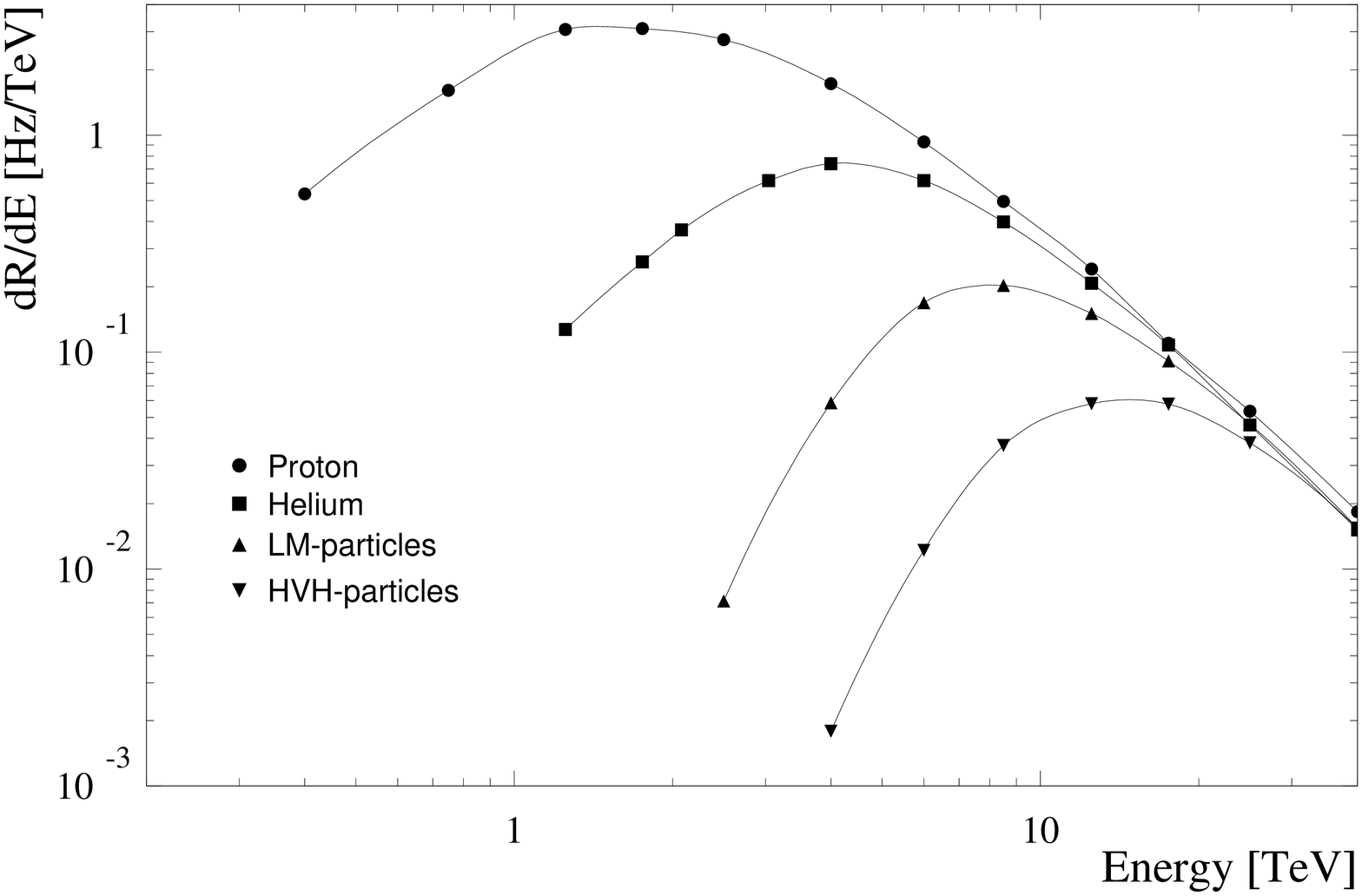,width=0.8\textwidth}}
  \end{center}
	 \protect\caption{Differential detection rates for different
	 nuclei according to individual spectra following an identical power
	 law. For a single telescope
	 trigger a $2{\rm NN}/271 > 10$ ph.e.~condition was applied. For the
	 System trigger a 2/4 coincidence was required. Already on the
	 trigger level, a clear suppression of heavier nuclei against protons
	 can be seen. At the energy threshold for protons, this
	 suppression amounts to at least a factor of
	 10.}\label{fig-altdrate}
      \end{figure}

	 A {\it first} suppression of the heavier nuclei occurs on the {\it
	 trigger level}. Figure \ref{fig-altdrate} shows the detection rates
	 for different particles, assuming for all nuclei a differential
	 spectrum ${{\rm d}F/{\rm d}E}$ $=$ $0.25\cdot E^{-2.7}$ ${\rm
	 s^{-1}\,sr^{-1}\,m^{-2}\, TeV^{-1}}$.  As can be seen, at 1.5 TeV,
	 the energy threshold for protons, heavier nuclei are suppressed by
	 more than one order of magnitude.
	
	 Note that apparently the suppression of heavier nuclei is best at
	 the trigger threshold.
	 Remarkably a similar suppression occurs {\it de facto} also
	 at higher energies by sorting the events into bins according
	 to their {\it reconstructed} energies.
	 Since at a given energy heavier particles produce a smaller
	 Cherenkov light density, their energy is estimated (see next
	 section) to be smaller by a factor $\eta$ with $\eta\,\approx\,3,
	 5, 6\,$ for Helium, Oxygen and Iron induced air showers,
	 respectively. In an energy bin centered at the reconstructed energy
	 $E$, protons of the mean true energy $E$ are contained, but also
	 heavier particles with the mean true energy $\eta\,\cdot$\,E. Since
	 the flux of all primary particles rapidly decreases with increasing
	 energy, to first order approximation according to dF/dE $\propto$
	 E$^{-2.7}$, heavier particles are suppressed by a factor
	 $\eta^{-2.7}$. This effect is slightly counterbalanced by a
	 relatively larger effective area for heavier particles at higher
	 energies ($>10$ TeV), due to the larger (although less intensive)
	 light pool. Detailed studies of the separation capabilities at
	 higher energies ($>10$ TeV) are still under way.
	
	 A further important suppression of heavier particles is achieved by
	 an analysis of the stereoscopic IACT images which mirror the
	 longitudinal and lateral shower development, described by the
	 Hillas-parameters \cite{HIL85}. Pixels with a small S/N-ratio are
	 excluded from the analysis, by computing the image parameters only
	 from the so called ``picture'' and ``boundary`` pixels \cite{PUN92}.
	 Picture-pixels are all pixels with an amplitude above the ``high
	 tailcut'' (here 6 photoelectrons). Boundary pixels are all pixels
	 with an amplitude below the ``high tailcut'' but above the ``low
	 tailcut'' (here 3 photoelectrons) which are neighbours of a
	 picture-pixel.
	
\begin{figure}[htbp]
	 \begin{center}
	    \mbox{\epsfig{file=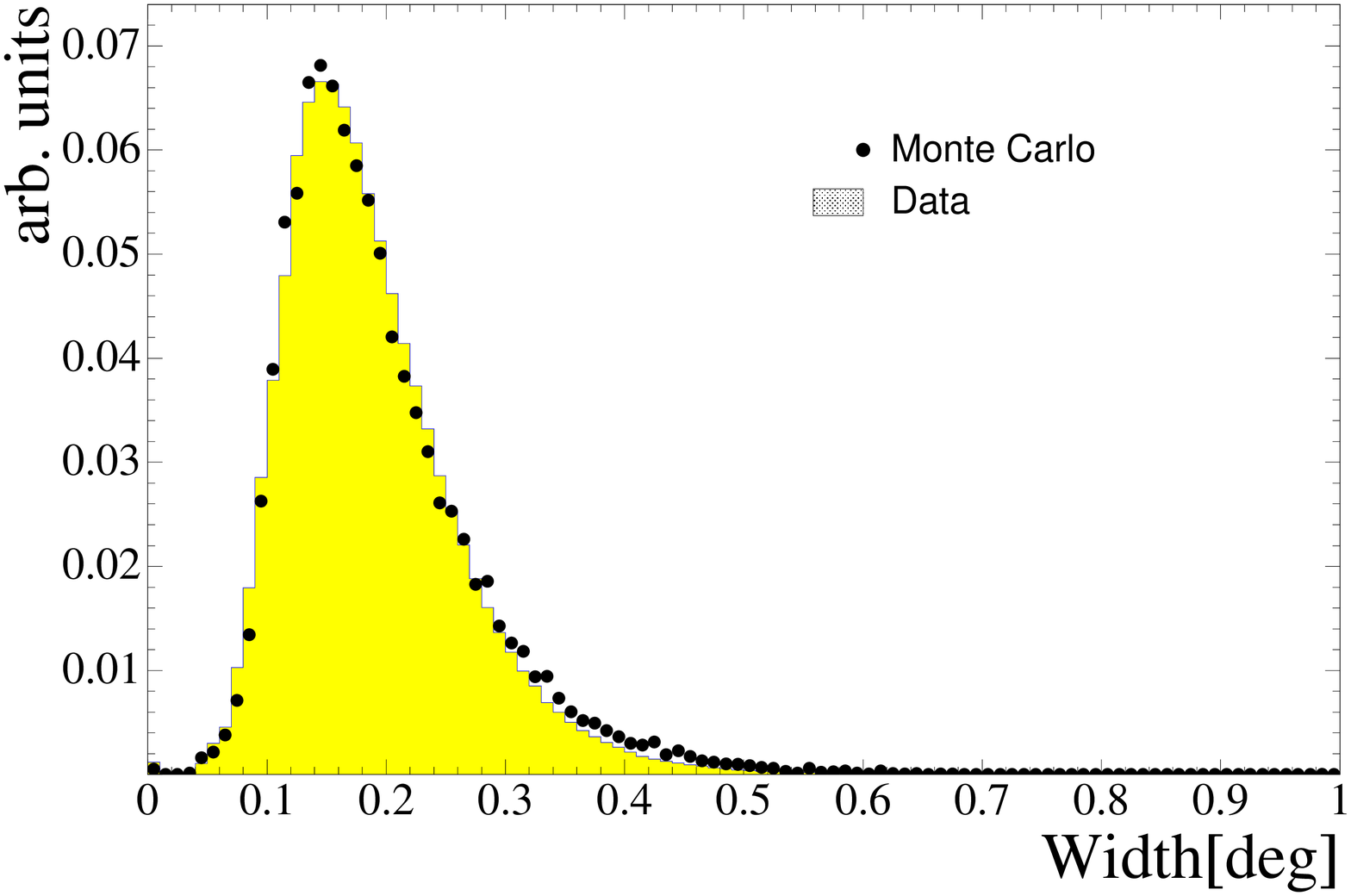,width=0.45\textwidth}}
	    \mbox{\epsfig{file=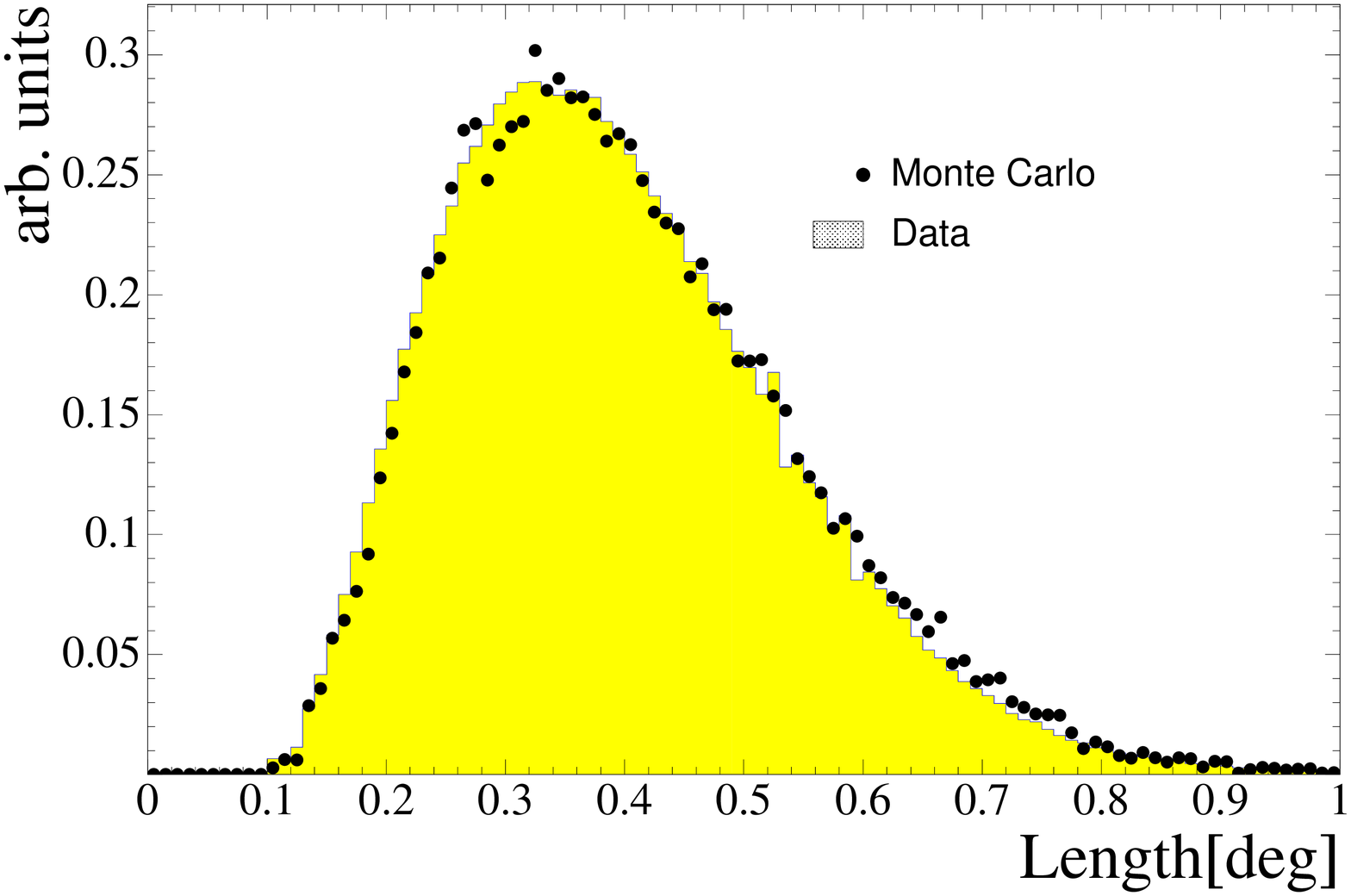,width=0.45\textwidth}}
	    \mbox{\epsfig{file=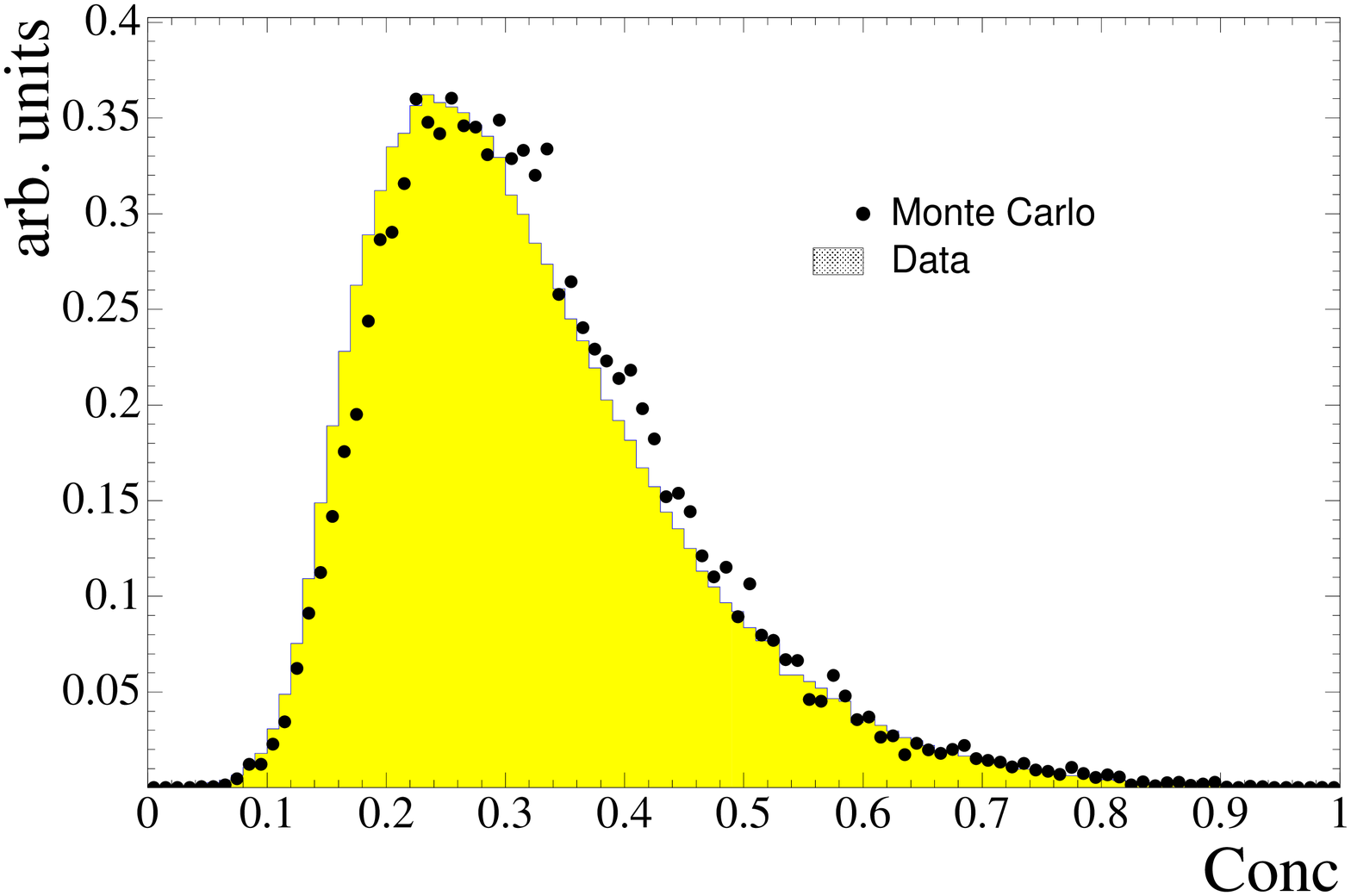,width=0.45\textwidth}}
	    \mbox{\epsfig{file=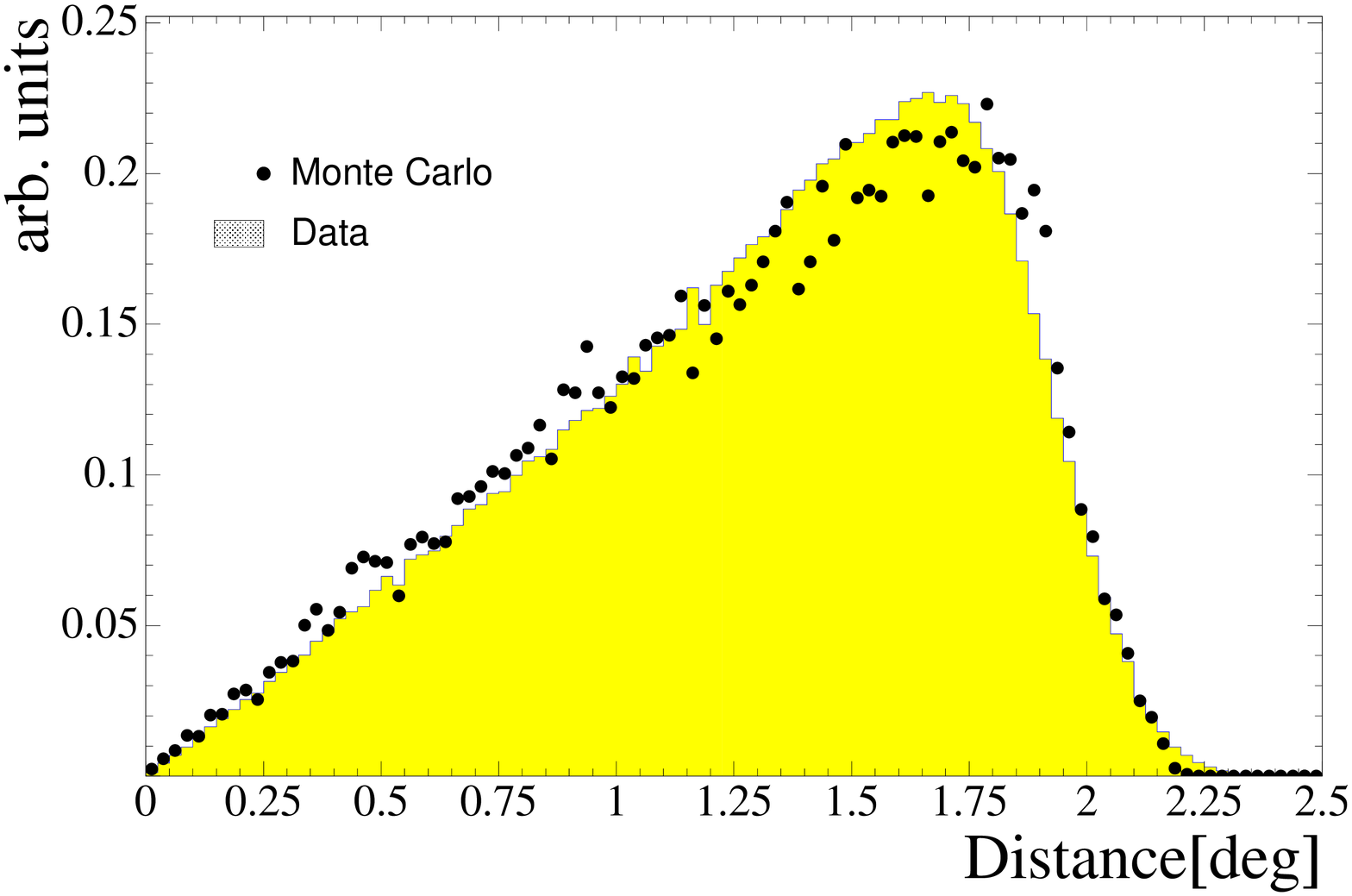,width=0.45\textwidth}}
	    \protect\caption{Comparison of Monte Carlo and measured image
	    parameters for cosmic rays for an assumed chemical composition
	    according to the compilation of \protect\cite{WIE94}.
	    A very good agreement between simulated and measured image
	    parameter distributions can be seen. Remind that identical cuts
	    where applied for the parameter calculations. \label{fig-images}}
	 \end{center}
      \end{figure}

	 Figure \ref{fig-images} shows the distribution of the most important
	 Hillas parameters for data and for Monte Carlo generated events.
	 Both in the data and the Monte Carlo events, a software threshold
	 has been applied, requiring two or more telescopes with at least two
	 pixels above 10 photoelectrons, and a sum {\sl ``Size''} of at least
	 40 photoelectrons recorded in the picture and boundary pixels.  The
	 Monte Carlo events have been weighted according to the chemical
	 composition from the literature (Table \ref{tab-wiebel}). The {\sl
	 Conc}-parameter, measuring the concentration of the amplitude in the
	 image, is defined as the amplitude in the two most prominent pixels
	 divided by the total amplitude in the image. Proton images are more
	 concentrated than images of heavier nuclei. The {\sl Distance}
	 represents the position of the image centroid in the camera. Since
	 hadronic showers fall in isotropic, the {\sl Distance} distribution
	 should rise linearly until it is cut by the edge of the camera. The
	 agreement between Monte Carlo and data image parameter distributions
	 in Figure \ref{fig-images} is very good.
	
	 Note that, since heavier particles are suppressed already on the
	 trigger level, the distributions in Figure \ref{fig-images} depend
	 only slightly on the assumed chemical composition.  In future work
	 we will try to use these small differences to determine the CR
	 chemical composition.
	
\begin{figure}[htbp]
	 \begin{center}
	    \mbox{\epsfig{file=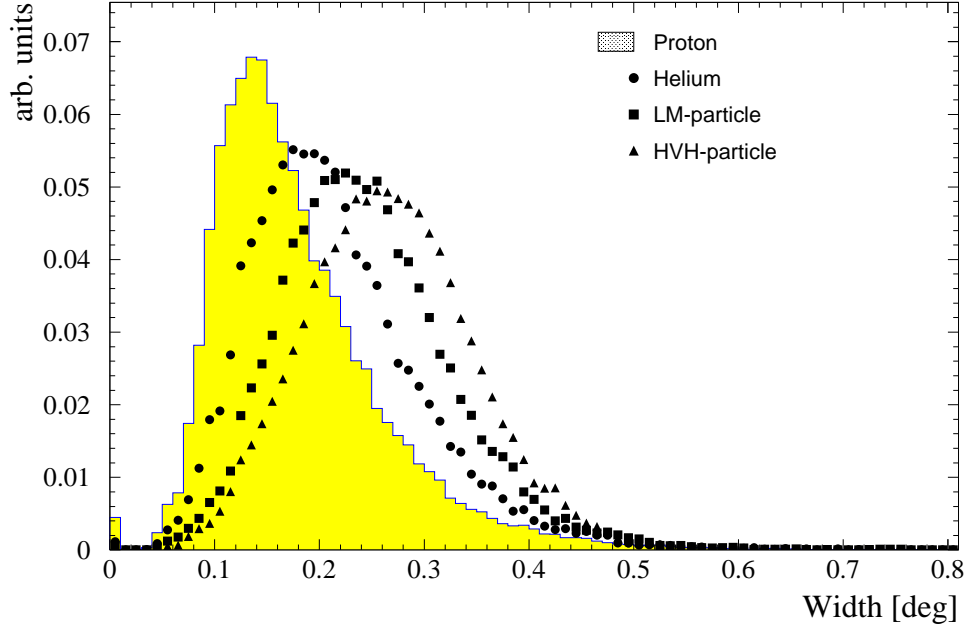,width=0.8\textwidth}}
	    \protect\caption{The {\sl Width}-distribution for the
	    particle groups from Table \protect{\ref{tab-wiebel}} after the
	    trigger condition $2{\rm NN}/271 > 10$ photoelectrons in each
	    telescope, requiring at least two triggered telescopes in each
	    event. The distributions are normalized to equal
	    area.\label{width-part}}
	 \end{center}
      \end{figure}

	 As outlined already in \cite{PLY98} and as seen in Figure
	 \ref{fig-img}, the {\sl Width} parameter, reflecting the lateral
	 extent of the air shower, is sensitive to the relatively larger
	 transverse momentum in showers induced by heavier particles and can
	 therefore be used to extract a data sample enriched with primaries
	 of a certain species.  Figure \ref{width-part} shows the
	 distributions of the {\sl Width}-parameter for the different
	 particle groups. The heavier the primary particle, the larger is the
	 {\sl Width} parameter.
	
	 In the following the parameter {\it mean scaled width}, introduced
	 in \cite{KON95} and first applied successfully to $\gamma$-ray data
	 in \cite{DAU97}, is used.
	 For each telescope $i$ the {\sl Width}-value
	 is normalized to the value expected for a proton shower
	 $\langle\,W(\textsl{Size}_i,r_i)\,\rangle_{\rm MC,p}$
	 given the sum of photoelectrons  of the image, {\sl Size$_i$},
	 and the distance $r_i$ of the shower core from the telescope.
	 The values obtained from the $n_{\rm tel}$
	 triggered telescopes are combined to the quantity
	 \begin{eqnarray}
	    W_{\rm scal} = 1/n_{\rm tel} \,
	    \sum_i^{n_{\rm tel}} W_i(\textsl{Size}_i, r_i)/
	    \langle\,W(\textsl{Size}_i,r_i)\,\rangle^p_{\rm MC}.
	 \end{eqnarray}
	
\begin{figure}[htbp]
	 \begin{center}
	    \mbox{\epsfig{file=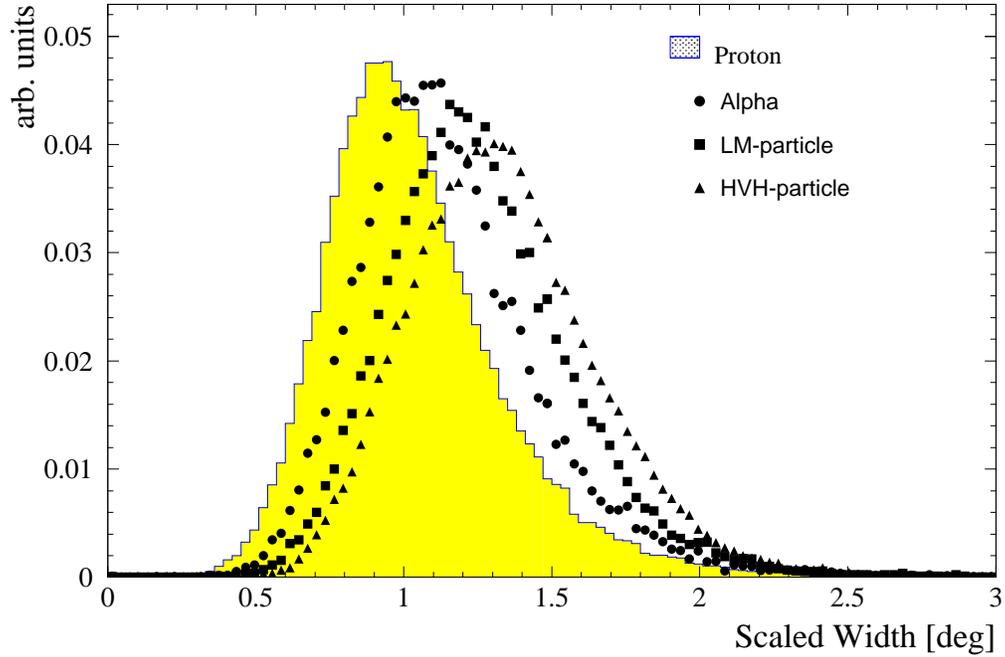,width=0.8\textwidth}}
	 \end{center}
	 \protect\caption{Scaled {\sl Width} parameter for the different
	 groups of nuclei (assuming a chemical composition from Table
\protect{\ref{tab-wiebel}}) as derived from the simulations (normalized to
	 equal area). The same trigger conditions as in Figure
	 \protect\ref{width-part} was applied.}\label{fig-scw}
      \end{figure}

\begin{figure}[htbp]
	 \begin{center}
	    \mbox{\epsfig{file=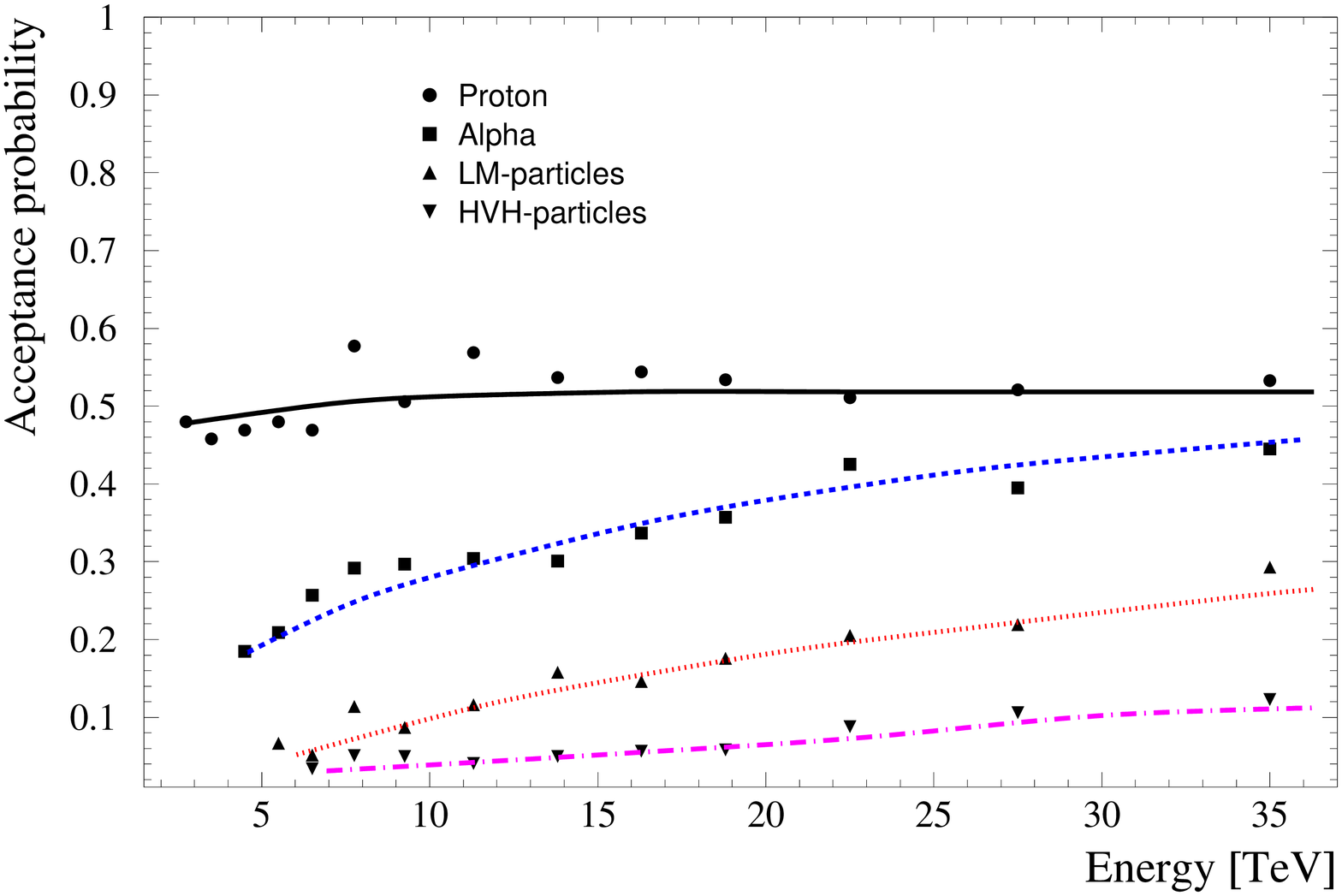,width=0.8\textwidth}}
	 \end{center}
	 \caption{Acceptance probability as a function of the energy
	 for a cut on the scaled {\sl Width} ($W < 0.85$) for different
	 nuclei. The lines are drawn to guide the eye.}\label{fig-acc}
      \end{figure}

	 Figure \ref{fig-scw} shows the distribution of the $W_{\rm
	 scal}$-parameter for the different groups of primaries (assuming a chemical
composition as given in Table \ref{tab-wiebel}). The $W_{\rm
	 scal}$-parameter, taking into account the distance and amplitude
	 dependence of the image width, allows one to enhance proton induced
	 showers among showers induced by all particles. The acceptances of
	 the different cuts are shown in Table \ref{tab-acc}. A cut in {\sl
	 Width$_{\rm scal}$} $<$ 0.85, for example, accepts $\sim 48$\% of the
	 primary protons, but only 20\% of the primary helium, and
	 $\simeq\,$10\% of the heavier nuclei.
	 The main advantage of scaling the {\sl Width}-parameter consists in
	 energy independent cut efficiencies for proton-induced air showers
	 and almost energy independent cut efficiencies for the heavier
	 primaries, as shown in Figure \ref{fig-acc}.
	 Since the image widths of proton-induced showers and
	 showers induced by heavier particles become more similar at higher
	 energies, the acceptance of heavier nuclei increases slightly with
	 their energy.
	
	 To summarize, at energies between 1 and 10 TeV the combined effect
	 of suppression of heavier nuclei by the detection principle and by
	 the image analysis enriches the data sample with proton-induced
	 showers by a large factor up to one order of magnitude, depending on
	 the used image shape cuts. In future we shall investigate if
	 additional image parameters can be used to obtain a similar
	 effective suppression also at energies above 10\,TeV. %
	 %***************************************************************************
      \subsection{Energy Determination}
\begin{figure}[htbp]
	 \begin{center}
	    \mbox{\epsfig{file=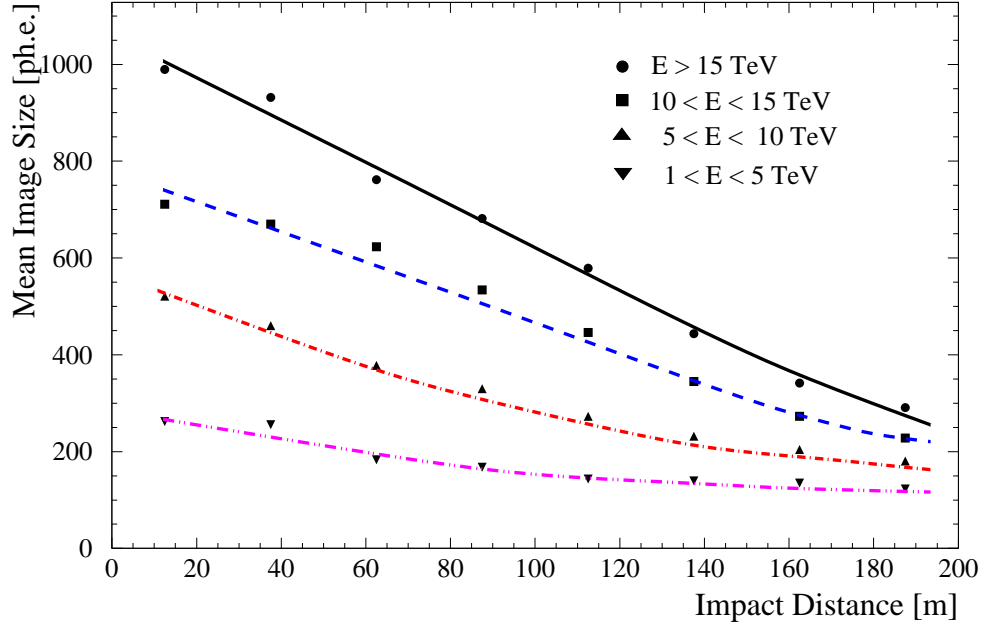,width=0.8\textwidth}}
	 \end{center}
	 \caption{Dependence of the image amplitude $S$ of the impact
	 distance $r$ for different primary energies $E$ for proton
	 showers.}\label{fig-fshape}
      \end{figure}

	 %***************************************************************************
	 For each triggered telescope, an energy estimate $E_i$ of the
	 primary particle is computed under the hypothesis of the primary
	 particle being a proton, from the image Size, $\textsl{Size}_i$,
	 measured in the $i$th telescope at the distance $r_i$ of the
	 telescope from the shower core. Averaging over all triggered
	 telescopes gives a common energy estimate. The energy estimate $E_i$
	 is determined by inversion of the relation $\textsl{Size}_i =
	 \langle\,\textsl{Size}(E,r)\,\rangle_{\rm MC,p}$ between primary
	 energy $E$, impact distance  $r_i$ and expected image Size
	 $\textsl{Size}_i$, as computed from the Monte Carlo simulations for
	 proton induced showers. For illustration purposes, the function
	 $\langle\,Size(E,r)\,\rangle_{\rm MC,p}$ is shown in
	 Figure \ref{fig-fshape} for 4 broad energy bins.
	 The expected number of photoelectrons decreases with increasing
	 distance from the shower core. The higher the proton's primary
	 energy is, the more pronounced is the light concentration near the
	 shower axis. More energetic showers penetrate more deeply into the
	 atmosphere. The tails of these showers give rise to the increased
	 light intensity near the shower axis in contrast to the flat light
	 pool of primary photons \cite{AHA98c}.

\begin{figure}[htbp]
	 \begin{center}
	    \mbox{\epsfig{file=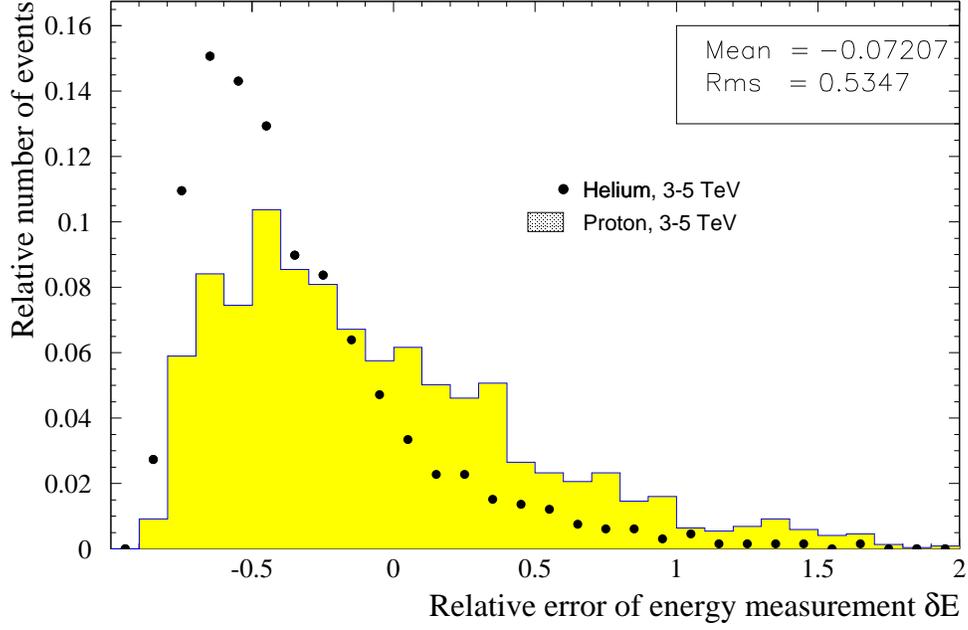,width=0.8\textwidth}}
	    \caption{Energy resolution of proton induced air showers with
	    an initial energy between 3 and 5 TeV. The distribution is
	    highly asymmetric. For an explanation see
	    text. The given values of Mean and rms error
	    relate to primary protons.\label{energy-res}}
	 \end{center}
      \end{figure}

	 This method leads to an energy resolution $\Delta E/E$ of
	 $\approx$~50\% for primary protons, as shown in Figure
	 \ref{energy-res} for proton induced showers. The energy resolution
	 is determined by the accuracy of the shower core reconstruction of
	 $\sigma_{r_i}\,=\,30$\,m and by the variations of the image size
	 (which is a function of $r_i$ and $E$). Cores which are
	 reconstructed too far away from the telescopes are partly
	 responsible for the long tail towards large values of $\Delta E/E$.
	 A second cause are the fluctuations of the image size due to
	 fluctuations in the shower development. As shown in Table
	 \ref{tab-energy} the energy resolution $\Delta E/E$ as a function of
	 primary proton energy is rather constant, an important requirement
	 for a robust and reliable deconvolution of the spectrum.

\begin{figure}[htbp]
	 \begin{center}
	    \mbox{\epsfig{file=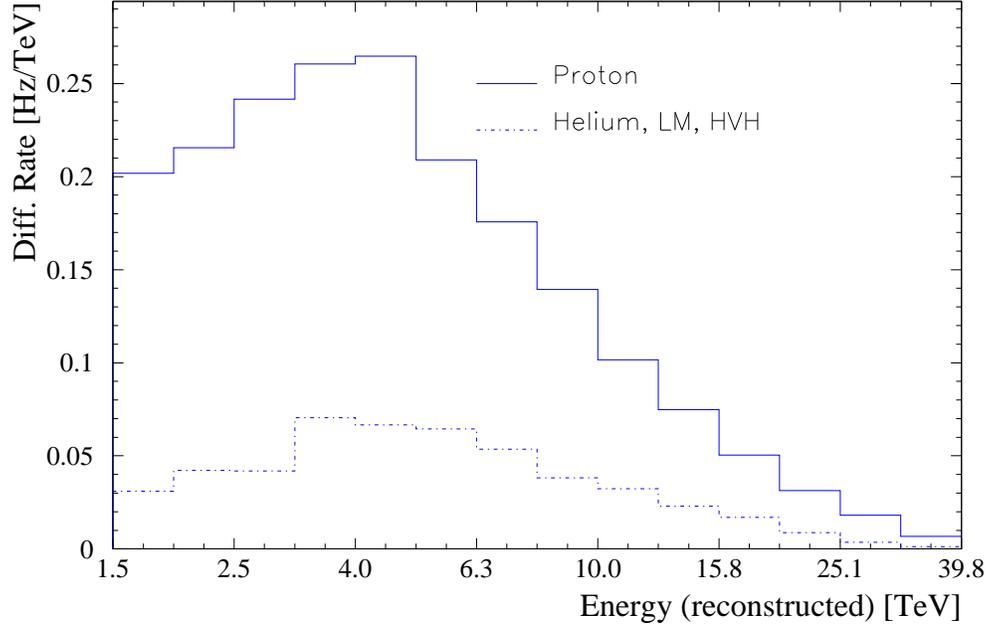,width=0.8\textwidth}}
	    \protect\caption{Differential detection rate of proton and the
	    group of helium, LM- and HVH-particles as function of the
	    reconstructed energy, for a cut on the scaled {\sl Width}
	    $<0.85$, assuming the chemical composition given in Table
	    \protect\ref{tab-wiebel}. \label{fig-end}}
	 \end{center}
      \end{figure}

	 The filled points in Figure \ref{energy-res} show the distribution
	 of $\Delta E/E$ for the helium-induced air showers.
	 As mentioned above, showers induced by heavier nuclei
	 produce, in comparison to proton-induced air showers
	 of the same energy, a lower Cherenkov light density at observation
	 level. This effect effectively suppresses, due to the steeply
	 falling spectra of CR primaries, the contamination of certain energy
	 bins with heavier nuclei.
	 In Figure \ref{fig-end} the differential detection rates
	 {\it after} the cut {\sl Width$_{\rm scal}\,<\,0.85$} are shown
	 as a function of the {\sl reconstructed} energy, assuming the
	 chemical composition from the literature (Table \ref{tab-wiebel}).
	 As can be seen, even if the helium to proton ratio would exceed the
	 value given in the literature by a factor of 2, the contamination of
	 the data sample by heavier particles is small, i.e.\ $<$~20\%,
	 taking into account also the cut efficiencies as given in Figure
	 \ref{fig-acc}.
	
	 %
	 %***************************************************************************
      \subsection{Method to determine the proton spectrum}
	 %***************************************************************************
	 %
	 The proton spectrum is determined using the standard method of
	 forward folding.

	 The Monte Carlo events of the particle group $i$
	 are weighted to correspond to a power law for the flux
	 dF/dE $=\,\alpha\,\cdot\,n_i$ E$^{-\gamma_i}$
	 where the $n_i$ and the $\gamma_i$ (except the $\gamma_i$ of the
	 proton component) reflect the chemical composition taken
	 from the literature (\cite{WIE94}, see Table \ref{tab-wiebel}).
	 The fitted parameters are the common scaling parameter $\alpha$ and
	 the spectral index of the proton component $\gamma_p$.
	 These two parameters are varied until the $\chi^2$-difference of the
	 observed histogram of reconstructed energies and the corresponding
	 histogram predicted with the weighted Monte Carlo events is
	 minimized. The fit is performed in the range from 1.5 to 3 TeV of
	 the reconstructed energy.
	
	 As we have shown in the previous sections, the contamination of the
	 data sample with heavier particles is small, especially in the
	 energy range from 1 to 3 TeV, and therefore the
	 result depends only slightly on their assumed abundances and
	 spectral index. This dependence has been studied in detail and will
	 be discussed in detail below.
	
	 Flux estimates at given energies are derived as follows:
	 Knowing the best fit value of the spectral index $\gamma_p$ of the
	 proton component, a correction function $U(E)$ is computed
	 from Monte Carlo simulations so that the differential flux of
	 protons at the reconstructed energy E can be
	 computed from the number n$_i$ of observed events in the $i$th
	 energy bin by
	 \begin{equation}
	    \rm dF/dE(E_i)\,=\,U(E_i)\,\cdot\,\frac{n_i}
	    {\Delta t \,\cdot\, \Delta E_i \,\cdot\, \kappa_p(E_i)
	    \,\cdot\, A_{\rm eff}(E_i)},
	    \label{diff}
	 \end{equation}
	 where $\Delta t$ is the observation time,
	 $\Delta E_i$ is the width of the $i$th energy bin,
	 $\kappa_p(E_i)$ is the acceptance for protons of the {\sl
	 Width$_{\rm scal}$}-cut, and $A_{\rm eff}(E_i)$ is the effective
	 area for proton registration. In this ansatz, the effect of the
	 energy resolution and the sample contamination by heavier particles
	 is accounted for by the function $U(E)$ which depends, for the
	 reasons mentioned above, only slightly on the assumed chemical
	 composition and on the {\sl Width$_{\rm scal}$}-cut in the energy
	 range of 1.5 to 3 TeV. Eq.\
	 \ref{diff} can strictly only be used if the proton spectrum indeed
	 follows the power law determined in the forward folding fit.
	 However, since the correction function $U(E)$ depends only weakly on
	 the spectral index, the method gives reasonable results, also for
	 spectra which deviate from the power law shape. %
	 %***************************************************************************
   \section{Results}
      \label{resu}
      %***************************************************************************
      \subsection{Data Set}
	 For the following analysis, the data primarily taken
	 for the observation of Mkn 501 during 1997 have been used.
	 Only runs taken under excellent weather and hardware
	 conditions were accepted. Table \ref{tab-data} gives a summary of
	 the data set.
	
	 The Mkn 501 data set was used because of its large fraction of small
	 zenith angle data. Furthermore, the solid angle region around
	 Mkn~501 does not contain very bright stars which cause excessive
	 additional noise. As a matter of fact, the strong $\gamma$-ray beam
	 from Mkn 501 in 1997 did not only supply informations of
	 astrophysical interest, but made it in addition possible to test the
	 simulation of electromagnetic showers and the simulation of the
	 detector response to these showers with unprecedented statistics
	 (in 1997, 38,000 photons were recorded).
	 The strong $\gamma$-ray beam could easily be excluded from the
	 analysis by rejecting all showers reconstructed within 0.3$^\circ$
	 from the source direction.
	
	 Identical cuts were applied to the measured data and the Monte Carlo
	 data. In addition to the cuts already mentioned above, a cut
	 on the distance $r$ of the shower axis from the central telescope of
	 $r\,<$ 175 m was applied. Only telescopes with a distance $r_i$
	 smaller than 200\,m from the shower axis entered the analysis,
	 suppressing by these means images close to the edge of the camera.
	 We apply a mean scaled width cut of {\sl Width$_{\rm
	 scal}\,<\,0.85$}, which minimizes, to our present understanding,
	 the systematic uncertainties caused by
	 the contamination of the data sample by heavier particles and by
	 the limited accuracy of the Monte Carlo simulations.
	
      \subsection{The proton spectrum}
	 The forward folding method described above gives a best power law
	 fit to the data in the energy range from 1.5 to 3 TeV for:
	 \begin{eqnarray}
	    {{\rm d}F\over{\rm d}E} \,=\,
	    (0.11 \pm 0.02_{\rm stat} \, \pm 0.05_{\rm sys})  \cdot
	    E^{-(2.72 \pm 0.02_{\rm stat} \, \pm 0.15_{\rm sys})}{\rm\
	    /\,s\,sr\,m^2\,TeV.}
	 \end{eqnarray}
	
\begin{figure}[htbp]
	 \begin{center}
	    \mbox{\epsfig{file=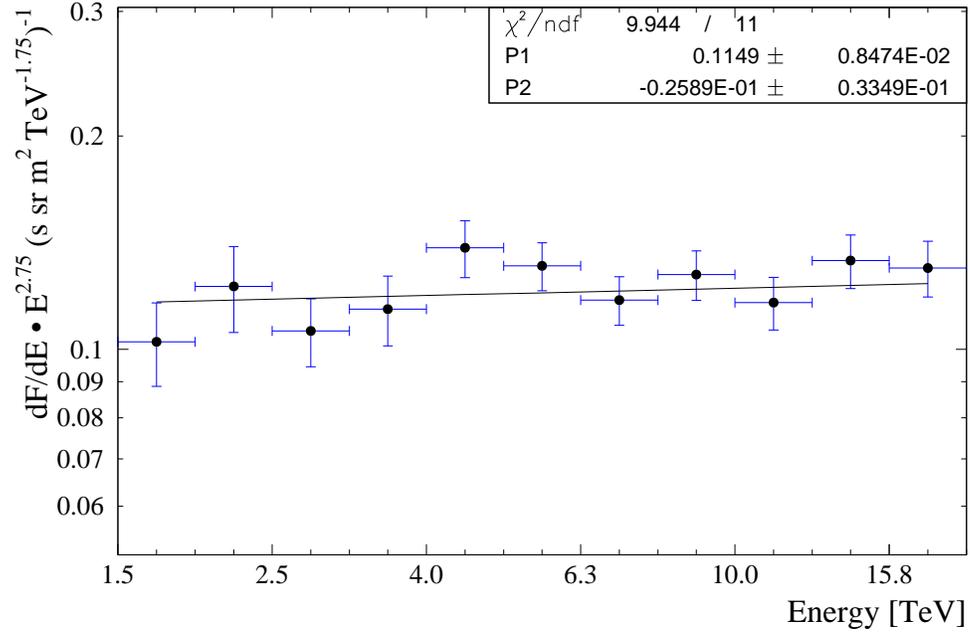,width=0.8\textwidth}}
	 \end{center}
	 \protect\caption{Differential energy spectrum of protons, obtained
	 using Eq.~\protect\ref{diff} and assuming the chemical composition
	 from Table \protect\ref{tab-wiebel}, multiplied by $E ^{2.75}$. The
	 cut in the scaled {\sl Width} was ${\textsl Width}_{\rm scal.} <
	 0.85$. Error bars are statistical only.}\label{fig-spec2}
      \end{figure}

	 In Figure \ref{fig-spec2} the differential
	 energy spectrum is shown assuming the chemical composition from
	 Table \ref{tab-wiebel}. This assumption allows to extend the energy
	 range of our measurement to energies above $>10$ TeV, as will be
	 explained later. As can be seen, a single power fits the data very
	 well. The systematic error on the spectral index is dominated by the
	 Monte Carlo dependence of the results and by the contamination of
	 the data sample by heavier particles.
	 The systematic error of the absolute flux is affected in addition by
	 an uncertainty in the energy scale of 15\%. We obtain an integral
	 flux above 1.5 TeV of $F(>1.5 {\rm\,TeV}) = 3.1 \pm 0.6_{\rm
	 stat.}\pm1.2_{\rm syst.} \cdot 10^{-2} /$ s sr m$^2$.

	 A rough estimate of the systematic errors can
	 be derived by varying the {\sl Width$_{\rm scal}$}-cut.
	 The different cuts lead to a varying percentage  of
	 heavier nuclei in the remaining data sample.
	 Table \ref{table-summary} summarizes the results for
	 cut-values between 1.15 and 0.65.
	 The derived spectral index varies between 2.68 and 2.76 and
	 the flux amplitude (differential flux at 1 TeV) varies between
	 0.08 and 0.13 ${\rm /s\,sr\,m^2\,TeV}$.

	 We have performed the following studies to estimate
	 the systematic error on the spectral index.
	
\begin{figure}[htbp]
	 \begin{center}
	    \mbox{\epsfig{file=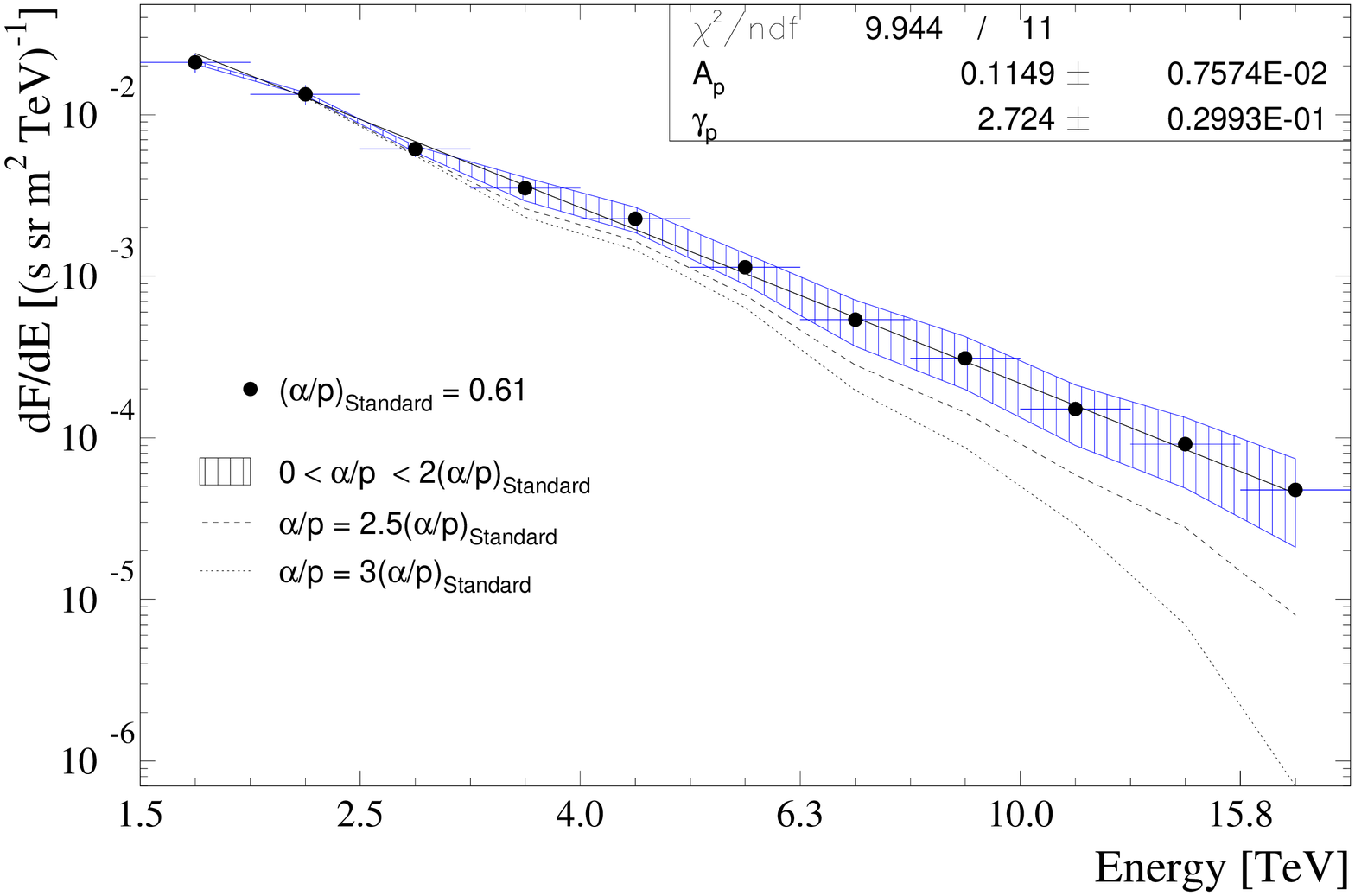,width=0.8\textwidth}}
	    \protect\caption{From experimental data reconstructed energy
	    spectrum of protons for a cut on the scaled {\sl Width} $<$ 0.85
	    and an assumed chemical composition according to
	    \protect\cite{WIE94} (black dots). The hatched area represents
	    the systematics connected with an over-estimation (no helium) and
	    under-estimation (doubled helium content) of the relative proton
	    content. Additional ratios are also given by the lines.
	    \label{sys_spectrum3}}
	 \end{center}
      \end{figure}

	 The dependence of the results on the assumed spectrum for
	 helium, LM- and HVH-particles was determined in a Monte Carlo study
	 by varying the
	 assumed abundance of the heavier particles over a wide margin.
	 Setting the assumed flux of one of the groups to zero or
	 increasing it by a factor of 2, yields the proton spectral indices
	 given in Table \ref{table-systematics}.
	 Since Helium has the lowest HEGRA energy threshold of the heavier
	 elements, the spectral index is most sensitive to the
	 abundance of the Helium component.
	 Setting the assumed Helium flux to zero results in a proton spectral
	 index of $\gamma_p\,=$ 2.79, and doubling it decreases the index to
	 2.73 from an assumed spectral index of $\gamma_p\,=$ 2.75.
	 (see also Figure \ref{sys_spectrum3}).

	 Table \ref{rates-comp} compares the measured CR detection rates with
	 the rates predicted by weighting the simulated events with the
	 CR spectra of Table \ref{tab-wiebel}. The rates are in very good
	 agreement. A much higher relative abundance of the heavier particles
	 than assumed in Table \ref{tab-wiebel} is therefore not probable.
	 Furthermore, if the Helium abundance would be much larger than
	 assumed here (more than two times the assumed abundance), the image
	 parameter distributions found in the data (see Figure \ref{fig-img}
	 and compare with Figure \ref{fig-images}) would not fit anymore the
	 Monte Carlo predictions.
	 Consequently, to our present understanding, the
	 systematic error in the spectral index due to the uncertainty in the
	 chemical composition is already estimated conservatively
	 by varying the relative abundances of the heavier elements
	 by factors between zero and two, and is in the order of 0.05.
	
	 Table \ref{last-wh} shows the results obtained from
	 experimental data under the extreme hypothesis of a pure CR proton
	 flux as function of the cut in {\sl Width$_{\rm scal}$}. Comparison
	 with Table \ref{table-summary} shows, that after applying tight cuts
	 ({\sl Width$_{\rm scal}\,<\,1.0$} or tighter) the results agree
	 nicely and consequently depend only weakly on the assumed chemical
	 composition.
	
	 The dependence of the spectral index on the detector performance and
	 on the atmospheric conditions has been derived as follows.
	 First, the data was divided into 4 parts of equal event statistics,
	 and the analysis was performed for each of the 4 subsamples. Second,
	 the data was divided into 4 seasonal parts and for each group the
	 spectrum was determined. The derived spectral indices are given in
	 Table \ref{tab-samples}. They are constant within $\sim0.05$.

	 The dependence of the spectral index on the details of the
	 Monte Carlo simulation (mainly threshold effects),
	 has been examined in the framework of determining the systematic
	 error on $\gamma$-ray spectra measured with the HEGRA IACT system.
	 The studies will be published elsewhere. The uncertainties
	 on the spectral index are currently estimated to be in the order of
	 0.1. The quadratic sum of these systematic errors (dependence on
	 assumed CR chemical composition 0.05, changing atmospheric and
	 detector conditions 0.05, threshold effects 0.1)
	 gives a total systematic error on the spectral index of 0.15.

	 Two effects dominate the systematic error on the flux amplitude.
	
	 The uncertainty in the energy scale of 15\% \cite{AHA98} translates
	 into an uncertainty of 30\% in the differential flux at a given
	 energy. The uncertainty of the differential flux in the energy range
	 from 1.5 to 3 TeV from threshold effects is estimated to be 10\%,
	 because with increasing energy the slope of the effective area
	 changes only slowly (compare with Figure \ref{fig-altcor}). Note
	 that since the energy threshold for heavier particles (Helium to
	 Iron) is much higher than for protons, the reconstructed proton flux
	 between 1.5 and 3 TeV is essentially independent of the assumed
	 contamination of the data sample by heavier particles. Figure
	 \ref{sys_spectrum3} shows the reconstructed proton spectrum varying
	 the Helium flux from zero to 3 times the value from the literature.
	 As can be seen, from 1.5 to 3 TeV, the reconstructed flux is to a
	 good approximation independent of the assumed Helium flux.
	
	 The quadratic sum of the systematic errors (energy scale 30\%,
	 threshold effects and cut efficiencies 15\%)
	 gives a total systematic of 35\%.

	 We also investigated wether broken power law models fit our data in
	 the energy range from 1 to 10 TeV better than single power law
	 spectra. Due to the limited energy resolution of $\Delta
	 E/E$\,=\,50\% for proton induced showers we would be able to detect
	 a break in the 1 to 10 TeV spectrum only for changes in the
	 differential index that are larger than $\sim 1$. The data do not
	 indicate such a break.
	
	 %
	 %***************************************************************************
	 %
   \section{Discussion}
      \label{disc}
      In this paper we used a new method to determine the Cosmic Ray
      proton spectrum in the energy range from 1.5 to 3 TeV with
      the stereoscopic IACT system of HEGRA.

\begin{figure}[htbp]
	 \begin{center}
	    \mbox{\epsfig{file=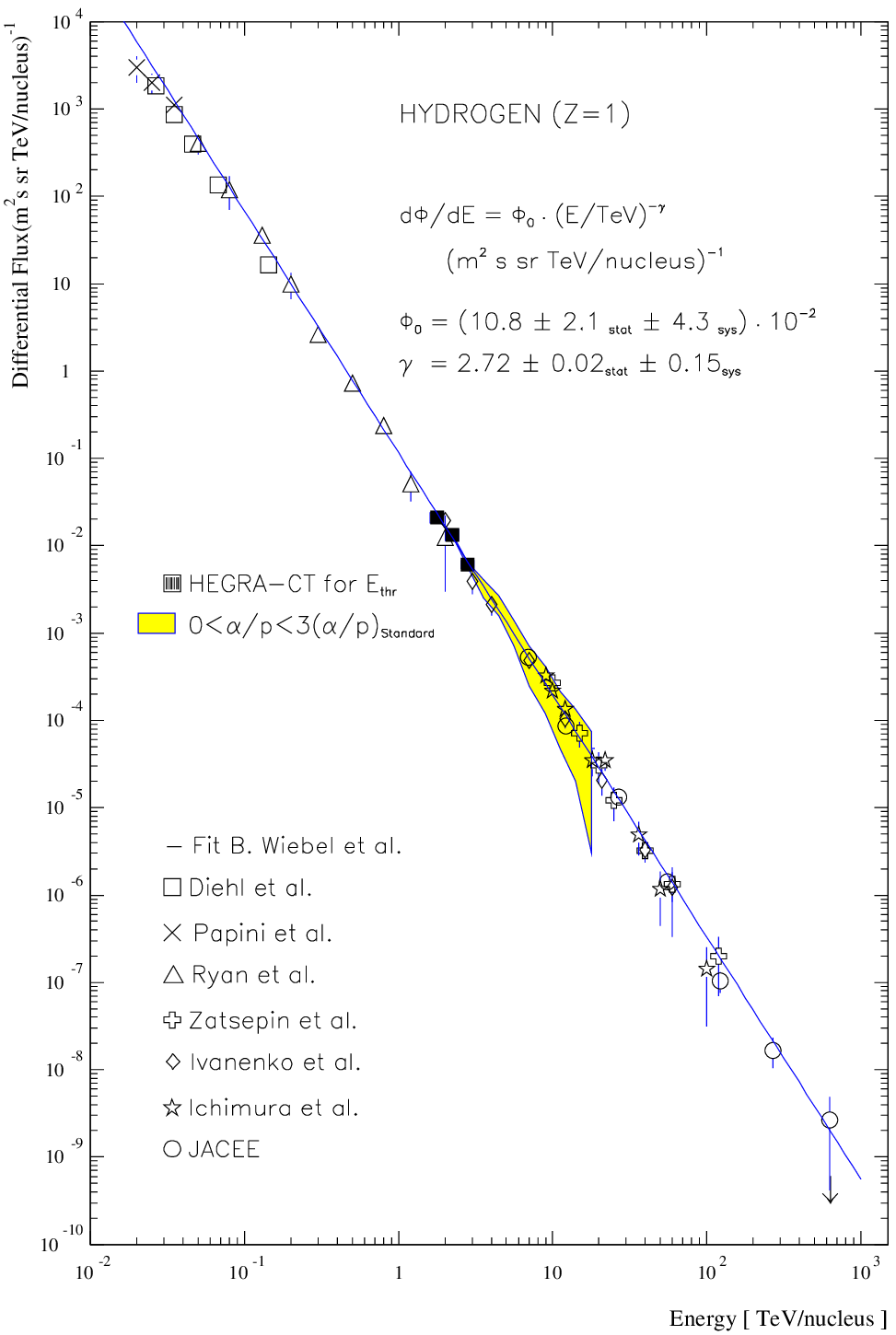,width=0.8\textwidth}}
	    \protect\caption{Comparison of our proton spectrum measurement
	    with other experiments. The black points are our measurements
	    around the threshold region of the HEGRA-CT-System. For
	    comparison also indicated are the results of previous
	    satellite and balloon-borne instruments. The shaded area
	    represents the systematic error of our
	    measurement caused by a variation of the
	    assumed $\alpha/{\rm p}$-ratio over the range
	    $0<\alpha/{\rm p}< 3(\alpha/{\rm p})_{\rm
	    Standard}$ relative to $(\alpha/{\rm p})_{\rm Standard} =
	    0.61$. The shaded area can be compared to the
	    extreme assumptions of Figure
	    \protect\ref{sys_spectrum3}.\label{proton}}
	 \end{center}
      \end{figure}

      As shown in Figure \ref{proton}, the results are in very good
      agreement with recent results of satellite and balloon-borne
      experiments (see the Figure for references).
      We have shown that the new technique yields a similar accuracy as
      achieved with the present day satellite and ballon-borne experiments,
      i.e.\ an error on the absolute flux of $\sim$50\% and an error on the
      spectral index of 0.15.

      Earlier claims about a possible cutoff in the proton spectrum at
      energies below 10 TeV are clearly not confirmed
      (e.g.~\cite{ZAT93,IVA93} and references therein).

      Our measurement of the proton spectrum is based on the large effective
      area of the atmospheric Cherenkov Technique of
      $\simeq\,3\cdot$10$^3$~m$^2$~sr for a field of view of $\simeq\,
      3$~msr, and the stereoscopic imaging technique which permits to
      reconstruct the protons' primary energy with the reasonable accuracy of
      $\Delta E/E$ of 50\%. The extraction of an almost pure proton data
      sample is possible due to a suppression of the number of heavier
      primaries by more than one order of magnitude using the multi-telescope
      trigger and the stereoscopic image analysis. The accuracy of the
      measurements is limited by an uncertainty in the energy scale of 15\%,
      by uncertainties of the detector acceptance, and by a residuum of
      heavier particles which could contaminate the data sample, if the
      relative abundance of heavier particles is much higher than presently
      believed. In future work we shall attempt to extend the measurement of
      the proton spectrum to higher energies. This might be possible by
      increasing the software threshold despite decreasing statistics.
      Improved cuts should also yield information about the spectrum of
      heavier nuclei.

      \acknowledgements
      The support of the German Ministry for Research and technology BMBF and
      of the Spanish Research Council CYCIT is gratefully acknowledged. We
      thank the Instituto de Astrofisica de Canarias (IAC) for supplying
      excellent working conditions at La~Palma. We gratefully acknowledge the
      work of the technical support staff of Heidelberg, Kiel, Munich and
      Yerevan. We thank H. Rebel and D. M\"uller for fruitful discussions.

      %%%%%%%%%%%%% Figures

      \newpage

      \newpage

      \newpage

      \newpage

      \newpage

      \newpage

      \newpage

      \newpage

      \newpage

      \newpage

      \newpage

      \newpage

      \newpage

      %%%%%%%% Tables

      \newpage

      \begin{table}[htbp]
	
	 \protect\caption{Parameters for the differential energy spectra
	 of different nuclei, taken from \protect\cite{WIE94},
	 using
	 \protect{
	 ${{\rm d}F\over{\rm d}E} = $
	 $\phi_0\cdot$
	 $E^{-\gamma}$
	 ${\rm/s\,sr\,m^2\, TeV}$.}
	 \protect\label{tab-wiebel}
	 }
	 %% \begin{center}
	 \begin{tabular}{|l|l|l|l|l|}\hline
	    Nucleus & p & He & LM & HVH \\\hline\hline
	    Atomic Number $A$ & 1 & 4 & 6-19 & 20-56 \\\hline
	    Mean Atomic Number $\langle A\rangle$ & 1 & 4 & 14 & 40
	    \\\hline
	    $\phi_0$ & 0.109$\pm$0.32 & 0.066$\pm$0.15 & 0.028$\pm$0.06 &
	    0.050$\pm$0.19
	    \\\hline
	    $\gamma$ & 2.75$\pm$0.02 & 2.62$\pm$0.02 & 2.67$\pm$0.02 &
	    2.61$\pm$0.03 \\\hline Proportion [at 1 TeV] & 0.43 & 0.26 & 0.11
	    & 0.20 \\\hline
	 \end{tabular}
	 %% \end{center}
	
      \end{table}

      \begin{table}[htbp]
	
	 \caption{Comparison of the integral rates for ALTAI (RSM) and
	 CORSIKA (HDPM) for the trigger $2{\rm NN}/271 >q_0$ ph.e.~and a
	 2/4 telescope coincidence.}\label{tab-altcor}
	
	 %% \begin{center}
	 \begin{tabular}{|l|l|l|l|l|}\hline
	    CR primary & p, CORSIKA & p, ALTAI & He, CORSIKA & He,
	    ALTAI\\\hline\hline $q_0$, ph.e. & R [Hz] & R [Hz] & R [Hz]
	    & R [Hz]\\\hline\hline 7  & 11.96 & 11.96 & 3.65 &
	    3.74\\\hline 10 & 6.63 &   6.61 & 2.14 & 2.19\\\hline
	    12 & 4.78 &   4.68 & 1.57 & 1.63\\\hline
	    15 & 3.62 &   3.56 & 1.21 & 1.28\\\hline
	    20 & 2.23 &   2.23 & 0.79 & 0.83\\\hline
	    30 & 1.22 &   1.22 & 0.46 & 0.47\\\hline
	 \end{tabular}
	 %% \end{center}
	
      \end{table}

      \begin{table}[htbp]
	
	 \caption{Acceptance probabilities for protons after
	 different scaled {\sl Width} cuts and the proportion for
	 different nuclei in the residual rate.}\label{tab-acc}
	 %% \begin{center}
	 \begin{tabular}{|l|l|l|l|l|l|l|}\hline
	    Cut, {\sl Width}$_{\rm scal}$ [deg] & 1.15 & 1.0 & 0.85 &
	    0.75 & 0.65 & 0.55\\\hline\hline Acceptance prob. for p &
	    0.854 & 0.706 & 0.477 & 0.293 & 0.143 & 0.049 \\\hline\hline
	    Proton proportion & 0.723 & 0.762 & 0.815 & 0.858 & 0.889 &
	    0.918\\\hline Helium proportion & 0.215 & 0.193 & 0.158 &
	    0.124 & 0.100 & 0.074\\\hline LM-proportion & 0.032 & 0.024 &
	    0.016 & 0.012 & 0.007 & 0.005\\\hline HVH-proportion & 0.031 &
	    0.021 & 0.011 & 0.007 & 0.004 & 0.003\\\hline
	 \end{tabular}
	 %% \end{center}
	
      \end{table}

      \begin{table}[htbp]
	
	 \caption{Energy resolution for proton induced air
	 shower.}\label{tab-energy}
	 %% \begin{center}
	 \begin{tabular}{|l|l|l|l|l|l|l|l|l|l|}\hline
	    Energy  [TeV] & 1.75 & 2.5 & 4.0 & 6.0 & 8.5 & 12.5 & 17.5 &
	    25.0 & 40.0
	    \\\hline\hline
	    $\delta E$ & 0.113 & 0.026 & -0.072 & -0.118 & -0.109 & -0.199
	    & -0.295 & -0.403 & -0.648 \\\hline
	    Resolution & 0.565 & 0.575 & 0.535 & 0.531 & 0.542 & 0.516 &
	    0.461 & 0.396 & 0.251 \\\hline
	 \end{tabular}
	 %% \end{center}
      \end{table}

      \begin{table}[htbp]
	 \caption{The data set.}\label{tab-data}
	 %% \begin{center}
	 \begin{tabular}{|l|l|}\hline
	    Runs & 79 \\\hline
	    Period & March-August 1997 \\\hline
	    max. $z$ [deg] & 20 \\\hline
	    $\bar{z}$ [deg] & 14.0 \\\hline
	    t, s & 191630 \\\hline
	    t, h & 53.2 \\\hline
	    Events & $\sim 2\cdot 10^6$ \\\hline
	    Events (e.g. $\textsl{Width}_{\rm scal} < 0.85$) & $\sim 6\cdot
	    10^5$\\\hline
	 \end{tabular}
	 %% \end{center}
	
      \end{table}

      \begin{table}[htbp]
	 \caption{Summary of proton spectrum for different scaled
	 {\sl Width} cuts, according to
	 \protect{${{\rm d}F\over{\rm d}E} = A_p\cdot E ^{-\gamma_p} \,
	 {\rm/s\,sr\,m^2\, TeV}$.}
	 \label{table-summary}}
	 %% \begin{center}
	 \begin{tabular}{|l|l|l|l|l|l|}
	    \hline
	    {\sl Width}$_{\rm scal}$ [deg] & 1.15 & 1.0 & 0.85 & 0.75
	    & 0.65\\\hline\hline
	    $A_p$ & 0.0829$\pm$0.0040 & 0.0975$\pm$0.0052 & 0.1149$\pm$0.0076
	    & 0.1206$\pm$0.0101 & 0.1274$\pm$0.0149 \\\hline
	    $\gamma_p$ & 2.675$\pm$0.022 & 2.709$\pm$0.024 & 2.726$\pm$0.030
	    & 2.726$\pm$0.038 & 2.758$\pm$0.053
	    \\\hline
	    \hline
	 \end{tabular}
	 %% \end{center}
      \end{table}

      \begin{table}[htbp]
	 \protect\caption{Reconstructed spectral indices of the proton
	 component with no or with doubled content of heavier particles
	 after according to the standard composition
	 \protect\cite{WIE94} calculated corrections. The assumed
	 proton spectral index was 2.75 ($W < 0.85$). This leads to a
	 systematic error of $\sim$ 0.04 due to an incorrectly assumed
	 chemical composition. \label{table-systematics}}
	 %% \begin{center}
	 \begin{tabular}{|l|l|l|l|}
	    \hline
	    Content & Helium & LM-particles & HVH-particles\\\hline
	    \hline
	    Double & 2.733 & 2.755 & 2.755 \\\hline
	    No & 2.793 & 2.762 & 2.762\\\hline
	 \end{tabular}
	 %% \end{center}
      \end{table}

      \begin{table}[htbp]
	 \protect\caption{Comparison of detection rates (given in [Hz]) of
	 the telescope system derived from Monte Carlo (with an assumed
	 chemical composition after \protect\cite{WIE94}), measurements and
	 data runs. The trigger condition was always 2 pixel above a
	 threshold $q_0$. NN signifies the next neighbour condition, MJ
	 the majority decision, which requires only two pixel not
	 necessarily neighboured for the trigger. The measured values come
	 from \protect\cite{BUL97}. The data values were derived directly
	 from Mkn501 data runs.} \label{rates-comp}
	 %% \begin{center}
	 \begin{tabular}{|l|l|l|l|l|l|l|l|l|}
	    \hline
	    System & Trigger & $q_0$, ph.e. $\to$ & 7 & 10 & 12 & 15 &
	    20 & 30 \\\hline\hline
	    & 2/4 & Measurement & 16.2 & 9.6 & 7.3 & 5.5 & 4.0 & 2.4 \\
	    & & Monte Carlo & 18.1 & 10.1 & 7.3 & 5.6 & 3.5 & 2.0 \\
	    NN & 3/4 & Measurement & 8.5 & 4.7 & 3.7 & 3.0 & 2.1 & 1.2
	    \\
	    & & Monte Carlo & 9.0 & 5.1 & 3.6 & 2.7 & 1.7 & 0.9 \\
	    & 4/4 & Measurement & 3.8 & 1.8 & 1.6 & 1.3 & 0.8 & 0.5 \\
	    & & Monte Carlo & 3.6 & 1.9 & 1.3 & 1.0 & 0.6 & 0.3
	    \\\hline & 2/4 & Measurement & 18.8 & 11.1 & 8.3 & 5.5 &
	    3.9 & 2.5 \\
	    & & Monte Carlo & 20.8 & 11.1 & 7.8 & 5.9 & 3.6 & 2.0 \\
	    & & Data Runs & & 9.1 & 7.7 & 5.9 & 3.9 & 2.2 \\
	    MJ & 3/4 & Measurement & 8.8 & 5.9 & 4.5 & 2.9 & 2.1 & 1.4
	    \\
	    & & Monte Carlo & 10.4 & 5.5 & 3.9 & 2.9 & 1.7 & 0.9 \\
	    & 4/4 & Measurement & 3.9 & 2.5 & 1.9 & 1.1 & 0.8 & 0.6 \\
	    & & Monte Carlo & 4.2 & 2.2 & 1.4 & 1.0 & 0.6 & 0.3
	    \\\hline
	 \end{tabular}
	 %% \end{center}
	
      \end{table}

      \begin{table}[htbp]
	 \caption{Summary of reconstructed indices from experimental data for
	 different scaled {\sl Width} cuts, assuming a pure proton sample in
	 the simulations. \label{last-wh}}
	 %% \begin{center}
	 \begin{tabular}{|l|l|l|l|l|l|}
	    \hline
	    {\sl Width}$_{\rm scal}$ [deg] & 1.15 & 1.0 & 0.85 & 0.75
	    & 0.65\\\hline\hline $A_p$ & 0.1216$\pm$0.0074 &
	    0.1322$\pm$0.0085 & 0.1397$\pm$0.0107 &
	    0.1538$\pm$0.0145 & 0.1552$\pm$0.0197 \\\hline
	    $\gamma_p$ & 2.647$\pm$0.028 & 2.676$\pm$0.030
	    & 2.690$\pm$0.035 & 2.727$\pm$0.044 & 2.756$\pm$0.059
	    \\\hline \hline
	 \end{tabular}
	 %% \end{center}
      \end{table}

      \begin{table}[htbp]
	
	 \caption{Reconstructed spectral index for different data
	 samples for a scaled {\sl Width} cut of
	 0.85.}\label{tab-samples}
	 %% \begin{center}
	 \begin{tabular}{|l|l|l|l|l|l|}\hline
	    Sample & 1 & 2 & 3 & 4 & $\Delta\gamma_{p,{\rm stat.}}$
	    \\\hline\hline Random & 13.6 h & 12.9 h& 13.6 h& 13.2 h&
	    \\\hline $\gamma_p$ & 2.73 & 2.73 & 2.72 & 2.71 & $\pm$
	    0.03  \\\hline\hline Periods &March-May & May & May-July &
	    July-August & \\\hline Observation time& 14.0 h & 13.8 h &
	    12.5 h & 11.8 h &  \\\hline $\gamma_p$ & 2.71 & 2.70 & 2.72
	    & 2.77 & $\pm$ 0.04 \\\hline
	 \end{tabular}
	 %% \end{center}
	
      \end{table}

      \newpage

      \bibliography{paper_re}

\end{document}

%% file: authors.tex
\author{
F. Aharonian\altaffilmark{1},
A.G. Akhperjanian\altaffilmark{7,1},
J.A.~Barrio\altaffilmark{3,2},
A.S. Belgarian\altaffilmark{7},
K.~Bernl\"ohr\altaffilmark{1,9},
J.J.G. Beteta\altaffilmark{3},
H. Bojahr\altaffilmark{6},
S. Bradbury\altaffilmark{2,8},
I. Calle\altaffilmark{3},
J.L. Contreras\altaffilmark{3},
J. Cortina\altaffilmark{3},
A. Daum\altaffilmark{1,11},
T. Deckers\altaffilmark{5},
S. Denninghoff\altaffilmark{2},
V. Fonseca\altaffilmark{3},
J.C. Gonzalez\altaffilmark{3},
G. Heinzelmann\altaffilmark{4},
M. Hemberger\altaffilmark{1,11},
G. Hermann\altaffilmark{1,13},
M. Hess\altaffilmark{1,11},
A. Heusler\altaffilmark{1},
W. Hofmann\altaffilmark{1},
H. Hohl\altaffilmark{6},
I. Holl\altaffilmark{2},
D. Horns\altaffilmark{4},
A. Ibarra\altaffilmark{3},
R. Kankanyan\altaffilmark{1,7},
M. Kestel\altaffilmark{2},
O. Kirstein\altaffilmark{5},
C. K\"ohler\altaffilmark{1},
A. Konopelko\altaffilmark{1,10},
H. Kornmeyer\altaffilmark{2},
D. Kranich\altaffilmark{2},
H. Krawczynski\altaffilmark{1,4},
H. Lampeitl\altaffilmark{1},
A. Lindner\altaffilmark{4},
E. Lorenz\altaffilmark{2},
N. Magnussen\altaffilmark{6},
H. Meyer\altaffilmark{6},
R. Mirzoyan\altaffilmark{2},
A. Moralejo\altaffilmark{3},
L. Padilla\altaffilmark{3},
M. Panter\altaffilmark{1},
D. Petry\altaffilmark{2,12},
R. Plaga\altaffilmark{2},
A. Plyasheshnikov\altaffilmark{1,10},
J. Prahl\altaffilmark{4},
C. Prosch\altaffilmark{2},
G. P\"uhlhofer\altaffilmark{1},
G. Rauterberg\altaffilmark{5},
C. Renault\altaffilmark{1},
W. Rhode\altaffilmark{6},
A. R\"ohring\altaffilmark{4},
V. Sahakian\altaffilmark{7},
M. Samorski\altaffilmark{5},
D. Schmele\altaffilmark{4},
F. Schr\"oder\altaffilmark{6},
W. Stamm\altaffilmark{5},
H.J. V\"olk\altaffilmark{1},
B. Wiebel-Sooth\altaffilmark{6},
C.A. Wiedner\altaffilmark{1},
M. Willmer\altaffilmark{5},
H. Wirth\altaffilmark{1}}